\documentclass{article}

\usepackage{titling}
\usepackage[centertags]{amsmath}
\usepackage{chicago,fancyhdr}
\usepackage{amssymb,amsfonts,eucal,oldgerm}
\usepackage[polutonikogreek,french,german,english]{babel}
\usepackage[colorlinks=true,citecolor=red,linkcolor=blue,urlcolor=blue,pdftex]{hyperref}
\usepackage[pdftex]{graphicx}
\usepackage[text={6in,8.25in},centering]{geometry}

\newcommand{\skipline}[1][1]{\vspace*{#1\baselineskip}}
\newcommand{\smidge}{\hspace*{.1em}}

\newtheorem{theorem}{Theorem}[subsection]

\newtheorem{defn}[theorem]{Definition}

\newtheorem{prop}[theorem]{Proposition}

\newcommand{\resetcounters}{%
  \setcounter{equation}{0}%
  \setcounter{theorem}{0}%
  \setcounter{figure}{0}%
}

\newcommand{\resetsec}{%
  \resetcounters%
  \renewcommand{\thetheorem}{\arabic{section}.\arabic{theorem}}%
  \renewcommand{\theequation}{\arabic{section}.\arabic{equation}}%
}

\newcommand{\english}{\selectlanguage{english}}

\newcommand{\french}[1]{\selectlanguage{french}#1\selectlanguage{english}}

\newcommand{\afourth}{\frac{1}{4}}

\title{On the Existence of Spacetime Structure\thanks{This paper is
    forthcoming in \emph{British Journal for Philosophy of Science},
    2015.}}

\author{Erik Curiel\thanks{I owe a great debt to Howard Stein's papers
    ``Yes, but\ldots: Some Skeptical Remarks on Realism and
    Anti-Realism'' and ``Some Reflections on the Structure of Our
    Knowledge in Physics'', both of which inspired the paper's spirit.
    I am not sure whether Prof.~Stein would endorse the paper's
    methods.  I have hopes he would.  It is a pleasure to thank the
    Philosophy Department at Carnegie Mellon for tough questioning
    after a colloquium in which I presented an earlier version of this
    paper, and in particular Richard Scheines, Teddy Seidenfeld and
    Peter Spirtes for pushing me on the arguments of
    \S\ref{sec:limits}; I thank as well the philosophy of physics
    reading group at Irvine, and especially Jim Weatherall, for
    penetrating questions about \S\ref{sec:limits}.  The paper is much
    stronger for my attempts to address their skepticism.  I also
    thank the Fellows at the Center for Philosophy of Science
    (2008--2009) at Pitt and several graduate students in the History
    and Philosophy of Science Department and the Philosophy Department
    at Pitt for insightful questions after an informal presentation of
    the paper.  I am grateful to Jeremy Butterfield for graciously
    harsh comments on an early draft.  John Norton, much as I suspect
    he would like to, cannot escape my gratitude for conversations in
    which he effortlessly showed me how to present simply what I could
    see only as complex.  And I thank Howard Stein and David Malament,
    as always, for more than I can well say.  \textbf{Author's
      address}: Munich Center for Mathematical Philosophy,
    Ludwigstra{\ss}e 31, Ludwig-Maximilians-Universit\"at, 80539
    M\"unchen, Germany; \textbf{email}:
    \href{mailto:erik@strangebeautiful.com}
    {\texttt{erik@strangebeautiful.com}}}}

\date{}

\begin{document}

\english

\nocite{stein-yes-but,stein94-struct-know}

\maketitle

\begin{quote}
  
  \begin{center}
    \textbf{ABSTRACT}
  \end{center}

  I examine the debate between substantivalists and relationalists
  about the ontological character of spacetime and conclude it is not
  well posed.  I argue that the so-called Hole Argument does not bear
  on the debate, because it provides no clear criterion to distinguish
  the positions.  I propose two such precise criteria and construct
  separate arguments based on each to yield contrary conclusions, one
  supportive of something like relationalism and the other of
  something like substantivalism.  The lesson is that one must fix an
  investigative context in order to make such criteria precise, but
  different investigative contexts yield inconsistent results.  I
  examine questions of existence about spacetime structures other than
  the spacetime manifold itself to argue that it is more fruitful to
  focus on pragmatic issues of physicality, a notion that lends itself
  to several different explications, all of philosophical interest,
  none privileged \emph{a priori} over any of the others.  I conclude
  by suggesting an extension of the lessons of my arguments to the
  broader debate between realists and instrumentalists.
\end{quote}

\newpage

\nocite{maxwell-attraction-origcite}
\begin{quote}
  
  [W]e must bear in mind that the scientific or science-producing
  value of the efforts made to answer these old standing questions is
  not to be measured by the prospect they afford us of ultimately
  obtaining a solution, but by their effect in stimulating men to a
  thorough investigation of nature.  To propose a scientific question
  presupposes scientific knowledge, and the questions which exercise
  men's minds in the present state of science may very likely be such
  that a little more knowledge would shew us that no answer is
  possible.  The scientific value of the question, How do bodies act
  on one another at a distance? is to be found in the stimulus it has
  given to investigations into the properties of the intervening
  medium.
  \begin{flushright}
    James Clerk Maxwell \\
    ``Attraction'', \emph{Encyclop{\ae}dia Brittanica} (9th ed.)
  \end{flushright}

\end{quote}

\skipline

\nocite{stein-yes-but}
\begin{quote}
  
  [B]etween a cogent and enlightened ``realism'' and a sophisticated
  ``instrumentalism'' there is no significant difference---no
  difference that \emph{makes} a difference.
  \begin{flushright}
    Howard Stein \\
    ``Yes, but\ldots---Some Skeptical Remarks on Realism and
    Anti-Realism''
  \end{flushright}

\end{quote}

\skipline

\tableofcontents

\thispagestyle{plain}

\section{Introduction}
\label{sec:intro}

\resetsec

The revival of the debate in recent years in the broader community of
philosophers over the ontic status of spacetime can trace its roots,
in part, to its revival in the community of physicists.
\citeN{belot-gr-interp} and \citeN{belot-earman-presoc-qg}, for
instance, claim that philosophers ought to take the debate seriously
because many physicists do.  I do not think that fact suffices as a
good reason for philosophers to take the debate as interesting, much
less even well posed, and so enter into it.  The active work of
physicists on our best physical theories should provide the fodder for
the work of the philosopher of physics most of the time.  Sometimes,
however, the physicists are confused or just mistaken, and it is then
our job to try to help set matters straight.  I believe that is the
case here.\footnote{See
  Curiel~\citeyear{curiel-modesty,curiel-gr-needs-no-interp} for
  extensive arguments to this effect on closely related matters, and
  for a defence of this claim as a fruitful philosophical attitude.}

Other philosophers in recent work have taken inspiration from the
traditional debates themselves.  \citeN{maudlin-st-subs}, for
instance, after a \emph{pr\`ecis} of the debate in the 17th and 18th
centuries and Kant's attempt to sidestep it, concludes, ``[G]ranting
that the world is \emph{an sich} a spatiotemporal object, we must face
a fundamental problem: Are space and time entities in their own
right?''  In this paper, I dispute that ``must.''

A virtue of Maudlin's approach, which his work shares with that of
many other contemporary philosophers no matter their inspiration, is
the foundation of his arguments on the structures of our best physical
theories and the use of those structures to guide metaphysical
argument.  I think the method falls short, however, in so far as it
treats those structures in abstraction from their uses in actual
scientific enterprises, both theoretical and experimental.  This
lacuna leaves the debate merely formulaic, without real content, at
the mercy of clever sophistications without basis in real, empirically
grounded scientific knowledge in the fullest sense.

\citeN[p.~1]{stein94-struct-know} admirably sums up the situation as I
see it.  I quote him at length, as he says it better than I
could\label{pg:stein-struc-know}:
\begin{quote}
  [L]et me \ldots hazard a rough diagnosis of the reason why some
  things that are (in my view) true, important, and obvious tend to
  get lost sight of in our discussions.  I think ``lost sight of'' is
  the right phrase: it is a matter of perspective, of directions of
  looking and lines of sight.  As at an earlier time philosophy was
  affected by a disease of system-building---the \emph{\'esprit de{}
    syst\`eme} against which a revulsion set in toward the end of the
  last century---so it has (I believe) in our own time been affected
  by an excess of what might be called the \emph{\'esprit de{}
    technique}\ldots: a tendency both to concentrate on such matters
  of detail as allow of highly formal systematic treatment (which can
  lead to the neglect of important matters on which sensible even if
  vague things can be said), and (on the other hand), in treating
  matters of the latter sort, to subject them to quasi-technical
  elaboration beyond what, in the present state of knowledge, they can
  profitably bear.  [W]hat I have described can be characterized
  rather precisely as a species of scholasticism\ldots.  In so far as
  the word ``scholasticism,'' in its application to medieval thought,
  has a pejorative connotation, it refers to a tendency to develop
  sterile technicalities---characterized by ingenuity out of relation
  to fruitfulness; and to a tradition burdened by a large set of
  standard counterposed doctrines, with stores of arguments and
  counterarguments.  In such a tradition, philosophical discussion
  becomes something like a series of games of chess, in which moves
  are largely drawn from a familiar repertoire, with occasional
  strokes of originality---whose effect is to increase the repertoire
  of known plays.
\end{quote}
In the spirit of Stein's diagnosis, rather than something formally
sophisticated I'm going to propose something crude and simple: in
order to try to avoid the sort of sterility that purely formal
technical elaboration can lead to, we should look at the way that
spacetime structures are used in practice to model real systems in
order to try to make progress on issues closely related to those
treated in the standard debate.  For I do think that there are
important, deep questions that we can make progress on in the vicinity
of that debate, questions of the sort that Maxwell alludes to in the
passage I quoted as one of this paper's epigraphs.  As Maxwell
intimates, however, in order for such questions to be investigated
profitably, they must be such as to support and stimulate ``the
investigation of nature.''  And that, I submit, can be accomplished
only when the questions bear on scientific knowledge in all its
guises, as theoretical comprehension and understanding, as evidential
warrant and interpretative tool in the attempt to assimilate novel
experimental results, as technical and practical expertise in the
design and performance of experiments, and as facility in the bringing
together of theory and experiment in such a way that each may
fruitfully inform the other.

To that end, in this paper I will argue that the way to find the
philosophically and scientifically fruitful gold in the metaphysical
dross is to formulate and address the questions in a way that
explicitly makes contact with both the theoretical and the
experimental aspects of our best current knowledge about the kinds of
physical system at issue.  One way of trying to do that is to pose and
investigate the questions explicitly in the context of what I will
call an investigative framework: roughly speaking, a set of more or
less exactly articulated and fixed theoretical structures for the
modeling of physical systems, along with a family of experimental
practices and techniques suited to the investigation of the type of
systems the theoretical tools appropriately model, in the way the
theory actually models them.  Different investigative frameworks, as I
show by constructive example, provide different natural criteria with
which to render determinate content to the question of the ontic
status of spacetime, with none privileged \emph{sub specie
  {\ae}ternitatis} over any of the others.  Those different criteria
yield different answers to the question, suitably formulated in the
given frameworks.  This should not be surprising, I think.  After all,
different sorts of scientific investigations naturally assume and rely
on different relations between individual spacetime points and
metrical (and other forms of spatiotemporal) structure, and it is
those relations that are supposed to serve as the criteria for
existence of individual spacetime points; the mathematical formalism
of the theory does not by itself fix a univocal relation with clear
\emph{physical} significance between points of the spacetime manifold
and geometrical structures, both local and global ones, that live on
the manifold.  I therefore dispute not only the force of Maudlin's
``must,'' but even more the cogency of the demand itself, baldly
formulated.

I begin in \S\ref{sec:hole} with an examination of a popular argument,
the so-called Hole Argument, that seems to urge a form of
relationalism.  I do this for two reasons.  First, because advertence
to the argument has become something of a mannerism in the debate, it
must be confronted; I conclude that it has no bearing one way or
another on the issues the debate purports to address.  Second, I
discuss it because it yields a useful schema for the production of
concrete criteria in the terms of which one can try to explicate the
difference between substantivalists and relationalists, such as it is.
I use that schema---whether the identification of spacetime points
must depend on the prior stipulation of metrical structure---to frame
the argument of the subsequent two sections of the paper.  In each of
those two sections I make the schematic criterion concrete in the
context of a particular form of investigative framework so as to
construct two arguments with contrary conclusions, one in support of
something like relationalism and the other something like
substantivalism to show that one can make the debate concrete in any
of a number of precise, physically significant ways, none \emph{a
  priori} privileged over the others, and that those ways will not in
general agree in their consequences.\footnote{I do not know of anyone
  in the literature who adopts exactly the schematic criterion I
  propose to found my two arguments.  (Perhaps
  Hoefer~\citeyearNP{hoefer-meta-st-subst,hoefer-abs-vs-rel-st} comes
  the closest.)  I use it because I think it captures the flavor of
  the criteria that are often stipulated when one or the other
  position is being argued for or against, \emph{viz}., schematically
  speaking, that the question of the existence of spacetime points
  boils down to the relation of those points to some fixed, underlying
  geometrical structure, such as the metric.  (See,
  \emph{e}.\emph{g}., \citeNP{earman-world-enough},
  \citeNP{butterfield-hole-truth},
  Maudlin~\citeyearNP{maudlin-subs-st-ari-eins,maudlin-st-subs},
  \citeNP{rynasiewicz-lessons-hole}, \citeNP{belot-rehab-rel},
  \citeNP{dorato-subst-rel-struc-st-real},
  \citeNP{huggett-reg-acct-rel-st},
  Pooley~\citeyearNP{pooley-pts-parts-struc-real,pooley-subst-rel-approach-st},
  \citeNP{belot-geom-poss}.)  This is all I require for the overall
  argument of the paper.  I use this particular schema, moreover, as
  only one example of the sort of criterion one could with some
  justification rely on in this debate, not because I think it is
  canonical or privileged in some way, but because it is popular and
  has a lot to say for it \emph{prima facie}.  My hope is that showing
  how the debate breaks down when this particular criterion is used
  will, at the least, strongly suggest that it would similarly break
  down no matter what sort of purely formal criterion of that sort one
  used.
  DiSalle~\citeyear{disalle-dyn-indiscern-st,disalle-understand-st} is
  a notable example of a contemporary philosopher who takes an
  approach much more sympathetic to my own.  (See
  \citeNP{friedman-rvw-disalle} for a thoughtful discussion of
  DiSalle's work.)  Robert Geroch (in private conversation) is a
  notable example of a contemporary physicist who does so.
  \citeN{dorato-is-struc-real-rel-disguise} is an interesting case of
  a philosopher who agrees with me that the contemporary debate is not
  well posed, but thinks there is a best answer to a proper
  reformulation of the debate.  \citeN{rynasiewicz-abs-vs-rel-outmode}
  agrees with me that the contemporary debate is not well posed, but
  he uses arguments I would not endorse.}

The opposed arguments and contrary conclusions of
\S\S\ref{sec:limits}--\ref{sec:pointless}, in conjunction with the
dismissal of the Hole Argument, do not decisively refute the claim
that there is a single, canonical way to explicate the idea of a
spacetime point and so to enter into debate over the existence of such
a thing.  As I urge in \S\ref{sec:subs-rel}, they strongly suggest it
is a question best settled in the context of a particular form of
investigation.  The investigation itself in tandem with pragmatic
considerations and {\ae}sthetic predilections will guide the
investigator in settling the form of the question and so the search
for its answer.  For a given spacetime theory, and even a given model
within the theory, depending on one's purposes and the tools one
allows oneself, either one can treat spacetime points as entities and
individuate and identify them \emph{a priori}, or one can in any of a
number of ways construct spacetime points as factitious, convenient
pseudo-entities, as it were.  Nothing of intrinsic physical
significance hangs on the choice, and so \emph{a fortiori} science
cannot guide us if we attempt to choose \emph{sub specie
  {\ae}ternitatis} between the alternatives---such a choice must
become, if anything, an exercise in scholastic metaphysics only.

In \S\ref{sec:embarass}, I extend the discussion to a host of other
types of spacetime structure, such as Killing fields and topological
invariants.  The attempt to formulate criteria for the physicality of
such other structures adds weight to the conclusion that such
questions require concrete realization in the context of something
akin to real science in order to acquire substantive content.  I
conclude in \S\ref{sec:valedict} with a brief attempt to show that the
arguments of this paper ramify into the debate between realists and
instrumentalists more generally, by dint, in part, of the picture of
science the arguments implicitly rely on.  The overarching lesson I
draw is that metaphysical argumentation abstracted from the pragmatics
of the scientific enterprise as we know it---science as an actually
achieved state of knowledge and as an ongoing enterprise of
inquiry---is vain.  Very little of real substance can be learned about
the nature of the physical world by studying only theoretical
structures in isolation from how they hook up to experimental
knowledge in real scientific practice, as is the endemic practice in
the current debate.  In particular, tracking the alleged ontological
commitments of a theory based on an analysis of its formal structure
alone is not a viable approach to the issue, as we cannot know what
structures the theory provides have real physical significance, and
what sort of real physical significance they do have, unless we
understand how the theory is successfully applied in practice.

The constructions I found the arguments on require the use of advanced
mathematical machinery from the theory of general relativity.  The
format of the paper does not allow for an introduction to most of it.
(For the interested reader, \citeNP{wald-gr} or
\citeNP{malament-fnds-gr-ngt}, for example, contains comprehensive
coverage of all material required.)  I have tried to segregate it as
much as possible so that those who do not want to trudge through it
will not have to while still following the general argument.  For
those who do want to skip most of the technical material, I recommend
the following: in \S\ref{sec:hole}, ignore the sketch of the Hole
Argument (the second and third paragraphs of the section), but read
the rest; in \S\ref{sec:limits}, read the first two paragraphs and the
last one; in \S\ref{sec:pointless}, read the first two paragraphs
(including definition~\ref{def:pointless-criterion}), and the final
two paragraphs.  (The remainder of the paper should not pose strenuous
technical difficulties.)  This course will convey almost the entirety
of my argument, bar supportive details the technical material purports
to provide.

\section{The Hole Argument}
\label{sec:hole}

\resetsec

In recent times, several physicists and philosophers have construed
Einstein's infamous Hole Argument so as to place it at the heart of
questions about the ontic status of spacetime points.  Its lesson, so
claimed, is that one cannot identify spacetime points without reliance
on metrical structure, that there is no ``bare manifold of points'',
as it were, under the metric field.\footnote{See, \emph{e}.\emph{g}.,
  \citeN{belot-gr-interp} and \citeN{gaul-rovelli-loopqg-diffeoinv}.
  Einstein himself originally formulated the Hole Argument to
  highlight what he regarded as problems of indeterminism for any
  generally covariant theory.  See
  \citeN{einstein14-form-grund-allg-rel} and
  \citeN{einstein-grossmann-14-kovar-feldgleich} for two versions of
  the original argument, Norton
  \citeyear{norton-how-found-efe,norton-gen-covar} for historical and
  critical discussion, and \citeN{earman-norton-hole-arg} for the
  introduction of the argument to the contemporary philosophical
  debate.}  In the contemporary literature, the debate is often posed
thus: should the manifold $\mathcal{M}$ by itself or the ordered pair
$(\mathcal{M}, \, g_{ab})$ be properly construed as the represention
of ``physical spacetime''?

This, in brief, is the argument.  Fix a spacetime model $(\mathcal{M},
\, g_{ab})$.\footnote{I may seem to be biasing the argument already,
  by demanding that one fix both a manifold and a metric to fix a
  model of spacetime.  In fact, though, by ``model of spacetime''
  here, I explicitly mean ``manifold \emph{cum} metrical structure'',
  irrespective of how the debate over what really represents physical
  spacetime resolves itself, so there is no bias here.}  For ease of
exposition, we stipulate that the spacetime be globally hyperbolic,
and so possesses a global Cauchy surface, $\Sigma$.  (We could do
without this condition at the cost of unnecessary technical details.)
Say that we know the metric tensor on $\Sigma$ and on the entire
region of spacetime to its causal past, $J^- [\Sigma]$.  (Note that
$J^- [\Sigma]$ contains $\Sigma$.)  It is known that this forms a well
set Cauchy problem, and so there is a solution to the Einstein field
equation that uniquely extends $g_{ab}$ on $J^- [\Sigma]$ to a metric
tensor on all of $\mathcal{M}$, yielding the original spacetime we
fixed.\footnote{This is not, strictly speaking, accurate.  If no
  restrictions are placed on the form of the metric, then in general
  the initial-value problem is not well set.  Indeed, even a few known
  ``physical'' solutions to the Einstein field equation possess no
  well set initial-value formulation, for example those representing
  homogeneous dust and some types of perfect fluid.  (See,
  \emph{e}.\emph{g}., \citeNP{geroch-pde}.)  We can ignore these
  technicalities for our purposes, though it may raise a serious
  problem for those who worry about indeterminism in the theory, one
  which, to the best of my knowledge, has not been addressed in the
  literature.}  In particular, the solution to the Cauchy problem
fixes the metric on the region to the causal future of $\Sigma$, $J^+
[\Sigma]$.  Now, let $\phi$ be a diffeomorphism that is the identity
on $J^- [\Sigma]$ and smoothly becomes non-trivial on $J^+ [\Sigma] -
\Sigma$.  No matter what else one takes the significance of the
diffeomorphism invariance of general relativity to be, at a minimum it
must include the proposition that the application of a diffeomorphism
to a solution of the Einstein field equation yields another, possibly
distinct solution.  Apply $\phi$ to $g_{ab}$ (but not to $M$ itself);
this yields a seemingly different metric---a different ``physical
state of the gravitational field''---on $J^+ [\Sigma] - \Sigma$, in
the sense that the same points of $J^+ [\Sigma] - \Sigma$ now carry
(in general) a different value for the metric.  This is the crux of
the issue, that the diffeomorphism applied to the metric has yielded a
different tensor field in the sense that the same points of the
spacetime manifold now carry a different metric tensor than before.

We now face a dilemma, the argument continues
(\citeNP{earman-norton-hole-arg}): we can either hold that the
fixation of the metric on $J^- [\Sigma]$ does not determine the metric
on $J^+ [\Sigma] - \Sigma$, a radical form of seeming indeterminism,
or else we can conclude that spacetime points in some sense have no
identifiability or existence or what-have-you independent of the prior
fixation of the metric tensor.  The argument concludes that the denial
of the independent existence of spacetime points is the lesser of the
evils (or, depending on one's viewpoint, the greater of the
goods).\footnote{Though it does not seem to be recognized in the
  literature, there are two different versions of the argument used by
  different investigators.  The one I rehearse here can be thought of,
  in a sense, as a generalization of the other.  The more specialized
  form, which Einstein himself formulated and used, assumes that
  spacetime has a region of compact closure, the nominal hole, which
  is devoid of ponderable matter (\emph{i}.\emph{e}., in which the
  stress-energy tensor vanishes) though it itself is surrounded by a
  region of non-trivial stress-energy; the diffeomorphism is then
  stipulated to vanish everywhere except in the hole, and the argument
  goes more or less as in the general case, with the emendation that
  now it is the distribution of ponderable matter that does not
  suffice to fix the physical state of the gravitational field.
  (\citeNP{earman-world-enough}, for example, uses the more general
  argument, whereas \citeNP{stachel-mean-gen-covar-hole} uses the more
  specialized form.)  I think the specialized form of the argument
  introduces a dangerously misleading red herring, \emph{viz}.,
  physical differences between regions of spacetime with non-vanishing
  stress-energy and those without.  There seems to me no principled
  way within the context of the theory itself to distinguish between
  such regions in a way that bears on metaphysical or ontological
  issues.  One of the regions, that with stress-energy, has
  non-trivial Ricci curvature; the other does not, though it may have
  non-trivial Weyl curvature.  That difference by itself, the only one
  formulable strictly based on the theory, can tell us nothing in the
  abstract about the ontic status of the spacetime manifold.  The
  introduction of the difference seems rather to bespeak an old
  prejudice that material sources should suffice to determine the
  physical state of associated fields, but this is not true even in
  classical Maxwell theory.  Indeed, the issue seems much less of a
  problem in general relativity, for in the case of the Maxwell field
  we cannot determine a \emph{physically} unique solution without
  imposing boundary conditions; otherwise, we are always free to add a
  field with vanishing divergence and curl to a solution to yield
  another that will have different physical effects on charged bodies.
  In general relativity, one does not need to do anything of the sort
  to determine a physically unique solution, so long as the initial
  data is well behaved in the first place.  (See, \emph{e}.\emph{g}.,
  \citeNP[ch.~10, pp.~243--268]{wald-gr}.)}

I want to make a crude and simple proposal, for it seems to me that
the debate has lost sight of a crude and simple, and yet fundamentally
important, fact: just because the mathematical apparatus of a theory
appears to admit particular mathematical manipulations does not
\emph{eo ipso} mean that those manipulations admit of physically
significant interpretation, much less that those apparently
mathematical manipulations are even coherent in and of
themselves.\footnote{\citeN{weatherall-hole}, whose conclusions I
  endorse, argues vigorously that the sort of manipulation employed in
  the standard form of the Hole Argument does not make even
  mathematical sense.  For the sake of argument, however, I will
  assume here that it does.  (If one likes, one can take that
  assumption as being in the service of a \emph{reductio}.)}  One has
the mathematical structure of the theory; one is not free to do
whatever it is one wants with that formalism and then claim, with no
foundation in practice, that what one has done has physical
import.\footnote{\citeN[p.~149]{stachel-mean-gen-covar-hole} neatly
  describes the current attitude in the literature towards mathematics
  in physical theories:
\begin{quote}
  A current trend among some philosophers of science is toward what I
  will call ``the fetishism of mathematics.''  By this I mean the
  tendency to assume that all the mathematical elements introduced in
  the formalization of a physical theory must necessarily correspond
  to something meaningful in the physical theory and, even more, in
  the world that the physical theory purports to help us understand.
\end{quote}} Once one has the mathematical formalism in hand, one must
determine what one is allowed to do with it, ``allowed'' in the sense
that what one does respects the way that the formalism actually
represents physical systems.  A simple example will help explain what
I mean: adding 3-vectors representing spatial points in Newtonian
mechanics.  This shows the need for an investigative context for the
fixing of what counts as admissible manipulations of the mathematical
formalism, for as a physical operation adding spatial points makes no
sense---there is no sense to be had from the idea of linearly
superposing two different spatial points in Newtonian theory as a
representation of a physical state of affairs.  For the purposes of
computing factitious quantities such as the center of mass, however,
it does make sense, though, again, not as an operation that has a
physical correlate in the world.

General relativity, in its usual incarnation, is formulated with the
use of differential manifolds with pseudo-Riemannian metrics.  It does
not \emph{ipso facto} follow that every well formed mathematical
operation one can perform on a manifold with such a metric has
physical significance.  It arguably makes mathematical sense to apply
a diffeomorphism of the manifold to the metric only, and not to the
underlying manifold at the same time.  That fact by itself does not
imbue the operation with physical significance.  It is exactly
considerations such as the Hole Argument highlights that show how
diffeomorphisms ought to be applied to solutions of the Einstein field
equation so as to have physical significance.  When one applies a
diffeomorphism, one must apply it to both the manifold and the metric.
As I shall argue, no other procedure has physical content.\footnote{If
  one adopts a certain definition of a differential manifold,
  \emph{viz}., that it is an equivalence class of ``diffeomorphic
  presentations'', then one will say that the proposed operation does
  not make even purely mathematical sense.  (\citeNP{weatherall-hole}
  comes to the same conclusion, based on different, but related,
  arguments.)  $\mathbb{S}^2$, for example, can presented as a certain
  submanifold of $\mathbb{R}^3$, or as a certain submanifold of a
  17-dimensional hyperbolid, or simply as a manifold in its own right;
  $\mathbb{S}^2 \times \mathbb{R}^2$ can be presented, as here, as a
  direct product of manifolds, or as $\mathbb{R}^4$ with a line
  removed; and so on.  In this case, ``pushing tensors around on the
  manifold by a diffeomorphism without also pushing the points
  around'', as required by the Hole Argument, is not an unambiguous
  notion, for strictly speaking manifold points are defined only up to
  diffeomorphism in the first place.  I do in fact accept the
  definition of a differential manifold as an equivalence class, but I
  am trying to be as charitable as possible to the proponents of the
  debate and the arguments standardly deployed in its carrying out, so
  I am willing to grant for the sake of argument that the required
  manipulations make mathematical sense.  In any event, it is not only
  philosophers who explicitly attempt to manipulate manifolds and
  objects in them, in the context of general relativity, in the way
  the Hole Argument requires; see, \emph{e}.\emph{g}.,
  \citeN{pons-salisbury-time-gencovar-theors-komar-bergmann} for
  physicists explicitly doing so.}

The Hole Argument is obviated by the fact that the application of
$\phi$ to the manifold \emph{cum} metric results only in a different
presentation of the same intrinsic metrical structure.  All observers,
no matter which diffeomorphic presentation of the manifold \emph{cum}
metric they use in their respective models, will agree on what is of
intrinsic physical significance in the possible interaction of
physical systems.  (Are those two bodies in physical contact?  Is heat
flowing from this one to that or vice-versa?  Can a light-signal be
sent from this to that?  Is gravitational radiation present?  And so
on.)  There is no logical or physical contradiction in taking
different diffeomorphic presentations of the manifold \emph{cum}
metric each as the representation of the same physical structure.  One
must simply stipulate that, in the context of general relativity, the
application of a diffeomorphism to the metric is a \emph{physically}
well defined procedure only when one also applies it to the (given
presentation of the) manifold itself.  The worry about determinism
thus evaporates, doing away with the dilemma.  How one then goes on to
try to characterize the ontic nature of spacetime points, if that is
the sort of thing one is into, may be influenced by this restriction
on the applicability of diffeomorphisms to solutions of the Einstein
field equation, or it may not.  The point of fundamental importance is
that this restriction results from \emph{both} pragmatic and semantic
considerations about the way that one may employ the formal apparatus
the theory provides so as to respect how solutions to the Einstein
field equation represent physically possible spacetimes in
practice---how it is that the formal structures of the theory acquire
real physical meaning.

In sum, I do not see why the Hole Argument drives one to conclude that
one should or should not attribute some form of existence to spacetime
points independent of the metrical structure.  There is no logical or
physical contradiction, for example, in taking the image of a point
under the action of $\phi$ to be ``the same spacetime point'' as its
pre-image, as depicted in a different presentation of spacetime,
irrespective of metrical structure.  In this case, a spacetime point
would be something like an equivalence class of ordinary mathematical
points under the relation of being related by a diffeomorphism.  An
exact formulation that avoids having this idea collapse into
triviality---given any finite number of points on a manifold, there is
a diffeomorphism that maps those points onto any permutation of them,
which seems to leave one with a single equivalence class containing
all points---requires some refinement.  One could do something like
the following: a spacetime point is a physical entity that one can
uniquely, or at least adequately and reliably, individuate and
identify by what is of intrinsic physical significance at the physical
event that occupies it, no matter the diffeomorphic presentation of
the manifold of events; it is an entity, in other words, individuated
and identified by the equivalence class of physical events under
diffeomorphic presentation.\footnote{Such a characterization would not
  necessarily rely on metrical structure at a point since, in general,
  one needs to fix the physical state on an open neighborhood of a
  point in order to fix the metric structure at that point by way of
  the Einstein field equation; one cannot solve the Einstein field
  equation ``point by point'', as it were.  The easiest way to see
  this is to note the non-uniqueness of vacuum solutions.  This is
  intimately bound up with the fact that the value of the
  stress-energy tensor at a point does not determine the value of the
  Weyl tensor (conformal structure) at that point.}  If one wants to
respond that bare spacetime points \emph{per se} even with what are
tantamount to unique labels attached (if the spacetime is not overly
symmetric) are dependent on physical phenomena under this definition
and inobservable to boot, and so unnecessary in the formulation of
physical theory, so as to conclude that they have no independent
metaphysical existence of one sort or another, I would not necessarily
disagree, but neither should I think that one requires the Hole
Argument to make the point, for the game of the Hole Argument is that
one cannot identify spacetime points in the absence of metrical
structure.  One need not invoke or rely on metrical structure to make
the sort of identification I suggest, as I will show by construction
in \S\ref{sec:pointless}.

The basis for my rejection of the Hole Argument, that a proper
understanding of diffeomorphism invariance and the way to properly
implement it as a formal procedure vitiates it, rests on a deeper
point.  I think the most unproblematic and uncontroversial claim one
can make about diffeomorphic freedom is that it embodies an
irremediable mathematical arbitrariness in the apparatus provided by
general relativity for the modeling of physical systems: the choice of
the presentation of the spacetime manifold and metric one uses to
model a physical system is fixed only up to
diffeomorphism.\footnote{\citeN{einstein-aether} makes a closely
  related point himself: ``The fact that the general theory of
  relativity has no preferred space-time coordinates which stand in a
  determinate relation to the metric is more a characteristic of the
  mathematical form of the theory than of its physical content.''}
There are restrictions on how one can apply diffeomorphisms to
solutions in practice in order for that application to be consistent
with the physical content of the theory, and those restrictions may
have philosophical significance, but they may not as well.  By itself,
that there is arbitrariness tells us nothing of interest about the
theory.

A comparison is edifying.  Classical mechanics as embodied in either
Lagrangian or Hamiltonian mechanics has a similar arbitrariness,
slightly different in each formulation of the theory.  In Lagrangian
mechanics, one is free to choose the Lagrangian function itself on the
tangent bundle of configuration space up to the addition of a scalar
field derived from a closed 1-form on configuration space (or, in more
traditional terms, up to the addition of a total time-derivative of a
function of configuration coordinates) without changing the family of
solutions the Lagrangian determines.\footnote{See, \emph{e}.\emph{g}.,
  \citeN{curiel-geom-ele}.}  In Hamiltonian mechanics, one is free to
choose any symplectomorphism between the space of states and the
cotangent bundle of configuration space, \emph{i}.\emph{e}., one may
choose, up to symplectomorphism, any presentation of phase space (or,
in more traditional terms, any complete set of canonical coordinates),
without changing the family of solutions the Hamiltonian function
determines.\footnote{\emph{Op.\ cit}.}  One feels no lack of
understanding of Lagrangian mechanics, no lacuna in its conceptual
resources, merely because one is free to choose the form of the
Lagrangian with wide latitude; just so, in Hamiltonian mechanics one
is not driven to investigate the ontic status of points in phase space
or of the physical quantities whose values one uses to label those
points, which ones get nominated `configuration' and which `momentum',
merely because one is free to choose whatever symplectomorphism one
likes in its presentation.  Consider the fact that one can run an
argument analogous to the Hole Argument in the context of Hamiltonian
mechanics, substituting ``phase space'' for ``spacetime manifold'',
``symplectomorphism'' for ``diffeomorphism'' and ``symplectic
structure'' for ``metric''.  Does that show anything of intrinsic
physical significance?  No serious person would argue so.  And in this
case, it would be manifestly absurd to ``apply a symplectomorphism
only to the symplectic structure and not the underlying manifold'': in
general the underlying manifold is a cotangent bundle and the
symplectic structure is the canonical one on it; pushing the
symplectic structure around on its own will yield a new symplectic
structure that is \emph{not} the canonical one, and so one manifestly
unphysical for the purpose of formulating Hamilton's equation.

The choice of Lagrangian or the choice of symplectomorphism rests on
nothing more than pragmatic considerations of the type adumbrated by
\citeN{carnap-emp-sem-ont} in his discussion of the choice of a
linguistic framework for the investigation of philosophical and
physical problems.\footnote{This is not to say that I consider the
  choice of a Lagrangian or a symplectomorphic presentation of phase
  space to be the choice of a Carnapian linguistic framework, only
  that the sorts of considerations that go into each choice are
  similar.}  One chooses on the basis of nothing more than what puts
one at ease in any of a variety of ways, from pragmatic considerations
such as what will be simple or useful for a particular investigation,
to those based on historical custom and {\ae}sthetic predilection.  It
is clear that the existence of inevitable, more or less arbitrary,
non-physical elements in the presentation of the models of a theory by
itself does not require of one a decision on the ontic status of any
entities putatively designated by the mathematical structures of
either Lagrangian or Hamiltonian mechanics.  More to the point, it is
clear in these cases that the physical significance of the theory's
models is not masked or polluted by the unavoidable arbitrariness in
the details of their presentations.

In the same way, the diffeomorphic freedom in the presentation of
relativistic spacetimes does not \emph{ipso facto} require
philosophical elucidation, in so far as it in no way prevents us from
focusing on and investigating what is of true physical relevance in
systems that general relativity models, what one may think of as the
intrinsic physics of the systems, so long as one respects the
pragmatic conditions for the application of diffeomorphisms to
solutions.  It is neither formal relations nor substantive entities
that remain invariant when one applies a diffeomorphism to a
relativistic spacetime; it is the family of physical facts the
spacetime represents.  (This line of thought already strongly suggests
that the debate between substantivalists and relationalists is not
well posed.)  One may represent those facts in a language some of
whose primitive terms designate ``spacetime points'' or not.  Further,
one may want to restrict the attribution of existence to what has
intrinsic physical significance in the context of our best physical
theories.  Then again, one may not.  It is irrelevant to our capacity
to use them in profitable ways in science and, more important, to our
comprehension of those facts and our understanding of the role they
play in our broader attempts to comprehend the physical world.

In the end, however, the most serious problem I have with the Hole
Argument, and all other arguments analogous to it, comes to this:
nothing I can see militates in favor of taking the Hole Argument as
bearing on the ontic status of spacetime points, \emph{just because}
the Hole argument by itself provides no independent, clear and precise
criterion for what ``existence independent of metrical structure''
comes to.  That idea has no substantive content on its own.  In the
next two sections, I will show this by exhibiting two plausible,
precise criteria for what the idea may mean in the contexts of two
different types of investigation, which in the event lead respectively
to opposed conclusions.

\section{Limits of Spacetimes}
\label{sec:limits}

\resetsec

In this section, I propose an argument in favor of the view that one
cannot identify spacetime points in the absence of metrical structure,
and so, \emph{a fortiori}, that one cannot attribute to the spacetime
manifold any existence independent of that structure; the provision of
a precise criterion for the existence of spacetime structure, grounded
in both the structure and the application of physical theory, drives
the argument.  In the event, two criteria natural to the investigative
context will suggest themselves, a weaker one based on the idea of the
identifiability of spacetime points and a stronger one based on their
existence (in a precise sense).

\begin{figure}
 \centering
 \includegraphics{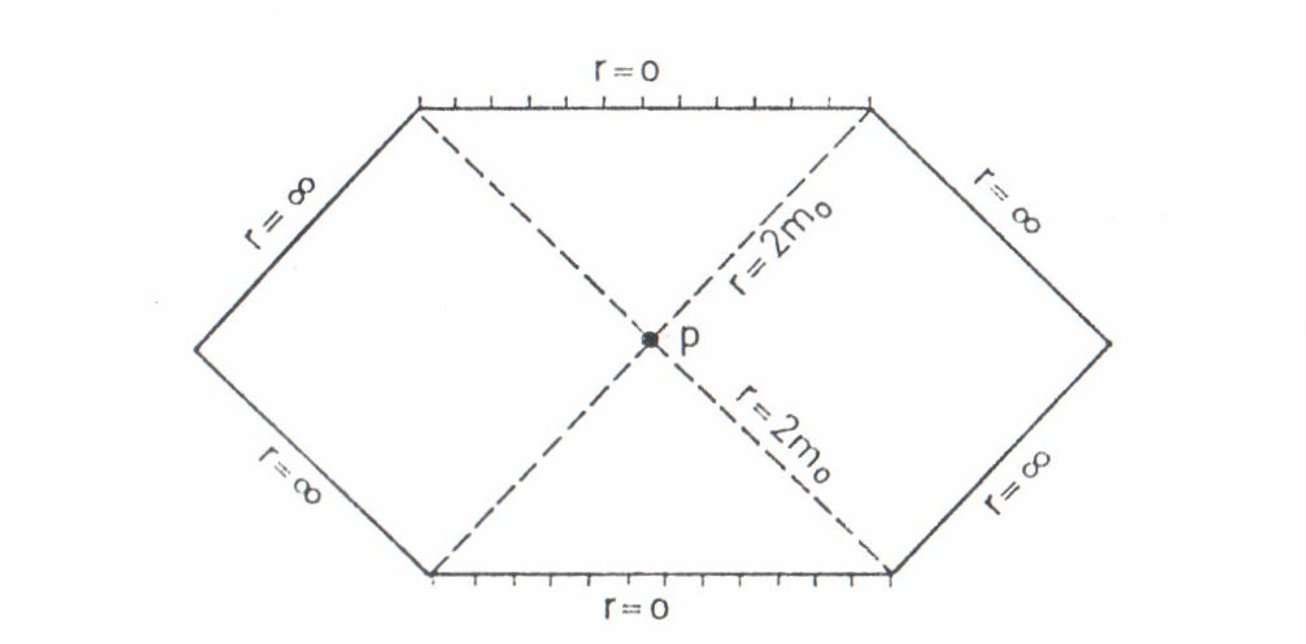}
 \caption{\label{fig:schwarz}Carter-Penrose diagram of Schwarzschild
   spacetime.  Each point in the diagram represents a 2-sphere in the
   spacetime manifold.  (This diagram is taken from \protect
   \citeNP{geroch-lim-sts}, with the author's permission.)}
\end{figure}

\begin{figure}
 \centering
 \includegraphics[height=4in]{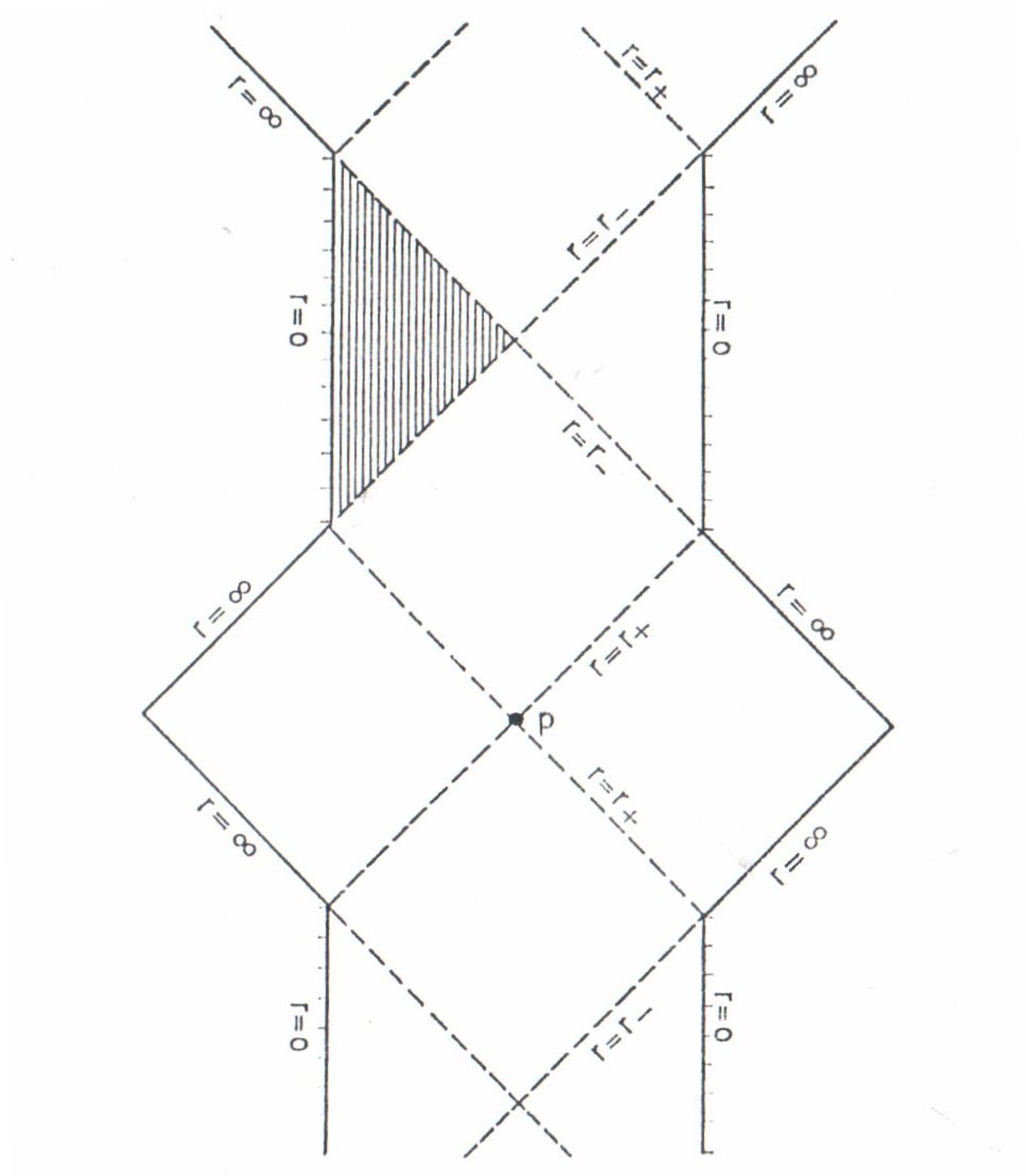}
 \caption{\label{fig:reiss-nord}Carter-Penrose diagram of
   Reissner-Nordstr\"om spacetime.  Each point in the diagram
   represents a 2-sphere in the spacetime manifold.  (This diagram is
   taken from \protect \citeNP{geroch-lim-sts}, with the author's
   permission.)}
\end{figure}

To treat a spacetime as the limit, in some sense, of an ancestral
family of continuously changing spacetimes is one of the ways of
embodying in the framework of general relativity two of the most
fundamental and indispensable tools in the physicist's workshop: the
idealization of a system by means of the suppression of complexity, so
as to render the system more tractable to investigation; and the
enrichment of a system's representation in a theory by the addition
(or reimposition) of complexity previously ignored (or ellided) in the
model the theory provides for the system.  As a general rule, the
fewer degrees of freedom a system has, the easier it becomes to study.
Schwarzschild spacetime (figure~\ref{fig:schwarz}) is far easier to
work with than Reissner-Nordstr\"om (figure~\ref{fig:reiss-nord}) in
large part because one ignores electric charge, and there is a natural
sense in which one can think of Schwarzschild spacetime as the limit
of Reissner-Nordstr\"om as the electric charge of the central black
hole decreases in magnitude to zero.\footnote{Schwarzschild spacetime
  is the unique spherically symmetric vacuum solution to the Einstein
  field equation (other than Minkowski spacetime); it represents a
  spacetime that is empty except for an electrically neutral,
  spherically symmetric, static central body or black hole of a fixed
  mass.  Reissner-Nordstr\"om is the generalization of Schwarzschild
  spacetime that allows the central structure to have an electric
  charge.  See, \emph{e}.\emph{g}., \citeN[ch.5,
  \S5]{hawking-ellis-lrg-scl-struc-st} for an exposition.}
Contrarily, as a general rule the more degrees of freedom one includes
in a system's model, the more phenomena that the system manifests the
model can represent, and with greater accuracy (or at least fineness
of detail).
A generic representation of such a limiting process can provide a
schema of both of these theoretical tools respectively, depending on
whether one enlarges or shrinks the number of degrees of freedom in
the limiting process.  As we will see, what in the idealized model one
may reasonably identify and attribute existence to may depend in
sensitive ways on the character of the more complex or simpler models
one starts from and the nature of the limiting process itself.  This
fact drives the argument I propose.  I will first discuss in some
detail two examples of such a limiting process in order to motivate
the two precise criteria I propose for the existence of spacetime
points independent of metrical structure.  

Before diving into the examples, however, I first characterize in the
abstract the limiting process itself.  I use the construction of
\citeN{geroch-lim-sts} (whose exposition I closely follow), which I
only sketch, to capture it.  (I simplify his construction in
non-essential ways for our purposes, and gloss over unnecessary
technicalities.)  Consider a 1-parameter family of relativistic
spacetimes, by which I mean a family $\{(\mathcal{M}_\lambda, \,
g^{ab} (\lambda))\}_{\lambda \in (0, 1]}$, where each
$(\mathcal{M}_\lambda, \, g^{ab} (\lambda))$ is a relativistic
spacetime with signature $(+, \, -, \, -, \, -)$ for $g^{ab}
(\lambda)$.  (It will be clear in a moment why I work with the
contravariant form of the metric tensor.)  In particular, I do not
assume that $\mathcal{M}_\lambda$ is diffeomorphic to
$\mathcal{M}_{\lambda'}$ for $\lambda \neq \lambda'$.  The problem is
to find a limit of this family, in some suitable sense, as $\lambda
\rightarrow 0$.  To solve the problem in full generality, we will use
a geometrical construction, gluing the manifolds $\mathcal{M}_\lambda$
of the family together to form a 5-dimensional manifold
$\mathfrak{M}$, so that each $\mathcal{M}_\lambda$ is itself a
4-dimensional submanifold of $\mathfrak{M}$ in such a way that the
collection of all of them foliate $\mathfrak{M}$.\footnote{In general
  what will result is not a foliation in the strict sense of
  differential topology, but will rather be a stratified space
  (\citeNP{thom-stratifies}).  It is close enough to a foliation,
  however, to warrant using the more familiar term for simplicity of
  exposition.}  $\lambda$ becomes a scalar field on $\mathfrak{M}$,
and the metrics $g^{ab} (\lambda)$ on each submanifold fit together to
form a tensor field $g^{AB}$ on $\mathfrak{M}$, of signature $(0, \,
+, \, -, \, -, \, -)$.  (I use majuscule indices for objects on
$\mathfrak{M}$.)  The gradient of $\lambda$ on $\mathfrak{M}$
determines the singular part of $g^{AB}$: $g^{AN} \nabla_N \lambda =
0$.  (This is why I work with the contravariant form of the metric;
otherwise, we could not contravect its five-dimensional parent in any
natural way with the gradient of $\lambda$.)  Note that $g^{AB}$ by
itself already determines the submanifolds $\mathcal{M}_\lambda$
(\emph{viz}., as the surfaces defined by $g^{AN} \nabla_N \lambda =
0$), and that it does so in a way that does not fix any identification
of points among them.  In other words, the structure I posit does not
allow one to say that a point in $\mathcal{M}_\lambda$ is ``the same
point in spacetime'' as a point in a different
$\mathcal{M}_{\lambda'}$ (as I shall discuss at some length below).

To define a limit of the family now reduces to the problem of the
attachment of a suitable boundary to $\mathfrak{M}$ ``at $\lambda =
0$''.  A \emph{limiting envelopment} for $\mathfrak{M}$, then, is an
ordered quadruplet $(\hat{\mathfrak{M}}, \, \hat{g}^{AB}, \,
\hat{\lambda}, \, \Psi)$, where $\hat{\mathfrak{M}}$ is a
5-dimensional manifold with paracompact, Hausdorff, connected and
non-trivial boundary $\partial \hat{\mathfrak{M}}$, $\hat{g}^{AB}$ a
tensor field on $\hat{\mathfrak{M}}$, $\hat{\lambda}$ a scalar field
on $\hat{\mathfrak{M}}$ taking values in $[0, \, 1]$, and $\Psi$ a
diffeomorphism of $\mathfrak{M}$ to the interior of
$\hat{\mathfrak{M}}$, all such that
\begin{enumerate}
    \item $\Psi$ takes $g^{AB}$ to $\hat{g}^{AB}$ (\emph{i}.\emph{e}.,
  $\Psi$ is an isometry) and takes $\lambda$ to $\hat{\lambda}$
    \item $\partial \hat{\mathfrak{M}}$ is the region defined by
  $\hat{\lambda} = 0$
    \item $\hat{g}^{AB}$ has signature $(0, \, +, \, -, \, -, \, -)$
  on $\partial \hat{\mathfrak{M}}$
\end{enumerate}
This makes precise the sense in which $\hat{\mathfrak{M}}$ represents
$\mathfrak{M}$ with a boundary attached in such a way that the metric
on the boundary ($\hat{g}^{AB}$ restricted to $\partial
\hat{\mathfrak{M}}$) can be naturally identified as a limit of the
metrics on the $\mathcal{M}_\lambda$ ($g^{AB}$ on $\mathfrak{M}$).  I
call $\{(\mathcal{M}_\lambda, \, g^{ab} (\lambda))\}_{\lambda \in (0,
  1]}$ an \emph{ancestral family} of the spacetime represented by
$\partial \hat{\mathfrak{M}}$, and I call $\partial
\hat{\mathfrak{M}}$ the \emph{limit space} of the family with respect
to the given envelopment.  In general, a given spacetime will have
many ancestral families, and an ancestral family will have many
different limit spaces.  For the sake of convenience I will often not
distinguish between $\mathfrak{M}$ and the interior of
$\hat{\mathfrak{M}}$.  (Although it is tempting also to abbreviate
`$\partial \hat{\mathfrak{M}}$' by `$\mathcal{M}_0$', I will not do
so, because part of the point of the construction is that different
spacetimes can have the same ancestral family.)

Before giving an example of the construction and putting it to work,
I discuss one of its features, that it parametrizes not only the
metrics but also the spacetime manifolds themselves.
\citeN[p.~181]{geroch-lim-sts} himself states in illuminating terms
the reason behind this.
\begin{quote}

  It might be asked at this point why we do not simply take the
  $g^{ab} (\lambda)$ as a 1-parameter family of metrics on a given
  fixed manifold $\mathcal{M}$.  Such a formulation would certainly
  simplify the problem: it amounts to a specification of when two
  points $p_\lambda \in \mathcal{M}_{\lambda'}$ and $p_{\lambda'} \in
  \mathcal{M}_\lambda$ ($\lambda \neq \lambda'$) are to be considered
  as representing ``the same point'' of $\mathcal{M}$.  It is not
  appropriate to provide this additional information, for it always
  involves singling out a particular limit, while we are interested in
  the general problem of finding all limits and studying their
  properties.

\end{quote}
To make the force of these remarks clear, consider the attempt to take
the limit of Schwarzschild spacetime as the central mass goes to 0.
In Schwarzschild coordinates, using the parameter $\lambda \equiv
M^{-1/3}$ (the inverse-third root of the Schwarzschild mass), the
metric takes the form
\begin{equation}
  \label{eq:schwarzschild}
  \left( 1 - \frac{2}{\lambda^3 r} \right) dt^2 - \left( 1 -
    \frac{2}{\lambda^3 r} \right)^{-1} dr^2 - r^2 (d\theta^2 + \sin^2
  \theta \smidge d\phi^2)
\end{equation}
This clearly has no well defined limit as $\lambda \rightarrow 0$.
Now, apply the coordinate transformation 
\[
\tilde{r} \equiv \lambda r, \quad \tilde{t} \equiv \lambda^{-1} t,
\quad \tilde{\rho} \equiv \lambda^{-1} \theta
\]
In these coordinates, the metric takes the form
\[
\left( \lambda^2 - \frac{2}{\tilde{r}} \right) d\tilde{t}^2 - \left(
  \lambda^2 - \frac{2}{\tilde{r}} \right)^{-1} d\tilde{r}^2 -
\tilde{r}^2 (d\tilde{\rho}^2 + \lambda^{-2} \sin^2 (\lambda
\tilde{\rho}) \smidge d\phi^2)
\]
The limit $\lambda \rightarrow 0$ exists and yields
\[
-\frac{2}{\tilde{r}} d\tilde{t}^2 + \frac{\tilde{r}}{2} d\tilde{r}^2 -
\tilde{r}^2 (d\tilde{\rho}^2 + \tilde{\rho}^2 \smidge d\phi^2)
\]
a flat solution discovered by \citeN{kasner-geom-thms-efe}.  If
instead of that coordinate transformation we apply the following to
the original Schwarzschild form~\eqref{eq:schwarzschild},
\[
x \equiv r + \lambda^{-4}, \quad \rho \equiv \lambda^{-4} \theta
\]
then the resulting form also has a well defined limit, which is the
Minkowski metric.  The two limiting processes yield different
spacetimes because it happens behind the scenes that ``the same points
of the underlying manifold get pushed around relative to each other in
different ways''.  Because the coordinate relations of initially
nearby points differ in different coordinate systems, those
differences get magnified in the limit, so that their final metrical
relations differ.  Thus, the limits in the different coordinates yield
different metrics.\footnote{\citeN{paiva-et-lims-sts} discuss in some
  detail an interesting class of different limiting spacetimes one can
  induce from Schwarzschild spacetime by taking the limit as the mass
  goes to zero and to infinity in different ways.  See
  \citeN{bengtsson-et-classics-lims-sts} for a similar discussion for
  Reissner-Nordstr\"om spacetime, as the electric charge and the mass
  respectively are taken to zero.}

In the language I introduced above, we should say that the difference
between the two limits consists in the different identifications each
makes among the points of different $\mathcal{M}_\lambda$.  That is
why it is inappropriate to work with a fixed manifold from the start.
To do so determines a unique limit, but we want to allow ourselves
different ways to take the limit, so that our ideal scientist can
ignore different facets of the complex system under study, and so
produce different idealized models of it.\footnote{Of course,
  sometimes is is appropriate for the scientist to take the limit of a
  family of metrics on a fixed background manifold.  An excellent
  example is in the statement and proof of the geodesic theorem of
  \citeN{ehlers-geroch-eom-small-bods-gr}.  In fact, they give an
  illuminating discussion of this very issue on p.~233.}  For example,
she may want to take the limit of Reissner-Nordstr\"om spacetime as
the mass goes to zero while leaving the electric charge fixed rather
than taking the limit as the electric charge vanishes, or she may want
to take the limit in a way that does not respect the spherical
symmetry of the initial system in order, \emph{e}.\emph{g}., to study
small perturbations of the original system.

To characterize the metrical structure of the limit space using
structure of members of the ancestral family, I introduce one more
construction.  An orthonormal tetrad $\xi (\lambda)$ at a point
$p_\lambda \in \mathcal{M}_\lambda$ is a collection of 4 tangent
vectors at the point mutually orthogonal with respect to $g_{ab}
(\lambda)$.  Let $\gamma$ be a smooth curve on $\mathfrak{M}$ nowhere
tangent to any $\mathcal{M}_\lambda$ such that it intersects each
exactly once.  $\gamma$ then is composed of a set of points $p_\lambda
\in \mathcal{M}_\lambda$, one for each $\lambda$.  A \emph{family of
  frames} along $\gamma$ is a family of orthonormal tetrads, one at
each point of the curve such that each vector in the tetrad is tangent
to its associated $\mathcal{M}_\lambda$, whose members vary smoothly
along it.  In general, a family of frames will have no well defined
limit in $\hat{\mathfrak{M}}$ as $\lambda \rightarrow 0$,
\emph{i}.\emph{e}., there will be no tetrad $\xi (0)$ at a point of
$\partial \hat{\mathfrak{M}}$ that the family $\xi (\lambda)$
converges to; in this case, I say the family is \emph{degenerate}.  It
is always possible, however, given a tetrad $\xi (0)$ at a point on
the boundary to find some family of frames that does converge to it.

Now, fix $\xi (0)$ at $p_0 \in \partial \hat{\mathfrak{M}}$ and a
family of frames $\xi (\lambda)$ that converges to it.  We can
represent the metric tensor $g_{ab} (\lambda)$ in a normal
neighborhood of $p_\lambda$ in $\mathcal{M}_\lambda$ using the normal
coordinate system that $\xi (\lambda)$ defines in the neighborhood.
In a normal neighborhood of $p_0$, the components of the metric with
respect to these coordinates converge as $\lambda \rightarrow 0$, and
the limiting numbers are just the components of $g_{ab} (0)$ at $p_0$
with respect to the normal coordinates that $\xi (0)$ defines.  In
this way, we can characterize all structure on the limit space based
on the behavior of the corresponding structures along the family of
frames in the ancestral family.

We are finally in a position to use this machinery to construct
concrete examples.  Consider a family $\{(\mathcal{M}_\lambda, \,
g^{ab} (\lambda))\}$ of Reissner-Nordstr\"om spacetimes each element
of the family having the same fixed value $M$ for its mass and all
parametrized by their respective electric charges $\lambda$, which
converge smoothly to 0.\footnote{I ignore the fact that electric
  charge is a discrete quantity in the real world, an appropriate
  idealization in this context.}  Construct their envelopment.  One
can now impose a natural collection of families of frames on the
family, with the limit space being Schwarzschild
spacetime.\footnote{The frames are natural in the sense that they
  conform to and respect the spherical and the timelike symmetries in
  all the spacetimes.  One could use this fact to explicate the claim
  that Schwarzschild spacetime is the canonical limit of
  Reissner-Nordstr\"om spacetime, in the sense that it is what one
  expects on physical grounds, whatever exactly that may come to, in
  the limit of vanishing charge while leaving all else about the
  spacetime fixed.}  Now, comparison of figures~\ref{fig:schwarz} and
\ref{fig:reiss-nord} suggests that something drastic happens in the
limit.  All the points in the throat of the Reissner-Nordstr\"om
spacetimes (the shaded region in the diagram) seem to get swallowed by
the central singularity in Schwarzschild spacetime---in some way or
other, they vanish.  Using our machinery we can make precise the
question of their behavior in the limit $\lambda \rightarrow 0$ in the
envelopment.

Consider the points in the shaded region in
figure~\ref{fig:reiss-nord}, between the lines $r = 0$ and $r = r^-$.
($r$ is the radial coordinate in a system that respects the
spacetime's spherical symmetry; the coordinate values $r^-$ and $r^+$
define boundaries of physical significance in the spacetime, which in
large part serve to characterize the central region of the spacetime
as a black hole.)  Fix a natural family of frames along a curve in
$\mathfrak{M}$ composed of points $q_\lambda$ each of which lies in
the shaded region in its respective spacetime.  It is straightforward
to verify that the family of frames along the curve does not have a
well defined limit: roughly speaking, the curve runs into the
Schwarzschild singularity at $r = 0$.  In this sense, no point in
Reissner-Nordstr\"om spacetime to the future of the horizon $r = r^-$
has a corresponding point in the limit space.  To sum up: one begins
with a family of Reissner-Nordstr\"om spacetimes continuously
parametrized by electric charge, which converges to 0, and constructs
the envelopment of the family; one constructs the limit space by a
choice of families of frames; the collection of families of frames
enforces an identification of points among different members of the
family of spacetimes, including a division of those points that have a
limit from those that do not; and that identification, in turn,
dictates the identification of spacetime points in the limit space
(which points in the ancestral family lie within the Schwarzschild
radius, \emph{e}.\emph{g}., and which do not).  Thus one can identify
points within the limit Schwarzschild spacetime, one's idealized
model, only by reference to the metrical structure of members of the
ancestral family; one can, moreover, identify points in the limit
space with points in the more complex, initial models one is
idealizing only by reference to the metrical structure of the members
of the ancestral family as well.  It is only by the latter
identification, however, that one can construe the limit space as an
idealized model of one's initial models, for the whole point is to
simplify the reckoning of the physical behavior of systems in
particular spatiotemporal regions of one's initial models, and most of
all at individual spacetime points of one's initial models.


One can, moreover, use different families of natural frames to
construct Schwarzschild spacetime from the same ancestral family, with
the result that in each case the same point of Schwarzschild spacetime
is identified with a different family of points in the ancestral
family.  More generally, different families of frames will yield limit
spaces different from Schwarzschild spacetime, with no canonical way
to identify a point in one limit space (one idealized model the
theoretician constructs) with one in another.  In other words, the
identification of points in the limit space depends sensitively on the
way the limit is taken, \emph{i}.\emph{e}., on the way the model is
constructed.  In consequence, in so far as one conceives of
Schwarzschild spacetime as an idealized model of a richer, more
complete representation, one can identify points in it only by
reference to the metrical structure of one of its ancestral families,
and one can do that in a variety of ways.

Now, say one wants to treat slightly aspherical, almost
Schwarzschildian spacetimes as a complexification of Minkowski
spacetime, in order to study how asphericities affect metrical
behavior.\footnote{One ought not confuse the idea of complexification
  I employ here---the making of a model more complex by the
  introduction of new representational structure---with the idea
  bandied about in other contexts in mathematical physics often also
  called `complexification', in which one takes a mathematical
  structure based on the real numbers and extends it to one based on
  the complex numbers.}  Because the limit spacetime will be almost
Schwarzschildian, its appropriate manifold is still $\mathbb{R}^2
\times \mathbb{S}^2$, the natural topology of Schwarzschild spacetime.
In this case, in one intuitive sense points will ``appear'', because
the topology of Minkowski spacetime is $\mathbb{R}^4$, so in some
sense one must ``compactify two topological dimensions'' to derive a
Schwarzschildian spacetime as a more complex limit.  There are many
ways to effect such a compactification; all the simplest, such as
Alexandrov compactification, work by the addition of an extra point or
set of points to the topological manifold to represent, intuitively
speaking, the bringing in of points at infinity to a manageable
distance from everything else.\footnote{See, \emph{e}.\emph{g}.,
  \citeN{kelley-gen-topo} for an account of methods of
  compactification, including the Alexandrov type.}  The difficulty of
these issues, however, is underscored by the fact that one can also
think of this as a case in which points rather \emph{disappear}:
$\mathbb{R}^2 \times \mathbb{S}^2$, after all, is homeomorphic to
$\mathbb{R}^4$ with a line removed!  Thus one could use an ancestral
family every member of which is $\mathbb{R}^4$ but that has as limit
space the manifold of Schwarzschild spacetime presented as the
manifold $\mathbb{R}^4$ with a line removed.\footnote{This is a
  concrete instance where thinking of two different diffeomorphic
  presentations of the same manifold---in this case, $\mathbb{R}^2
  \times \mathbb{S}^2$ and $\mathbb{R}^4$ with a line removed---as
  different manifolds leads to obvious difficulties, if not downright
  confusions.}

In this example, we will consider the attempt to introduce a central,
slightly aspherical body by physical construction in a Minkowskian
laboratory, as an experimentalist might do it.  For the sake of
concreteness, let us say that our experimentalist will, in his
representation of the experiment, use an Alexandrov compactification
of $\mathbb{R}^4$ to yield $\mathbb{R}^2 \times \mathbb{S}^2$ as the
presentation of the manifold of the limit space.  The physical
construction will proceed in infinitesimal stages, with a tiny portion
of matter introduced at each step distributed in a slightly aspherical
way (keeping, in an intuitive sense, the aspherical shape of the body
the same), and an allowance of a finite time to allow the ambient
metrical structure to settle down to an almost Schwarzschildian
character before the next step is initiated, until the central body's
mass reaches the desired amount.  (Intuitively, the finite time period
allows the metrical perturbations introduced by the movement of the
matter in and its distribution around the central body to radiate off
to infinity.)  One can represent this process with a limiting
ancestral family of Geroch's type in a more or less obvious way,
starting with Minkowski spacetime, \emph{viz}., the empty, flat
laboratory, and each member of the ancestral family representing the
laboratory at a particular stage of the construction, when a bit more
matter has been introduced and the perturbations have settled down.

Now, consider at the beginning of the process a small patch of space
in the laboratory not too far from the position where the central body
will be constructed.  We want to try to track, as it were, the
spacetime points in that patch during the enlargement of the central
body because we plan to investigate, say, how the metrical structure
in regions at that spatiotemporal remove from a central aspherical
body differ from each other for different masses of the central body.
(Because the Einstein field equation is nonlinear, and there is no
exact symmetry, one cannot just assume that slightly aspherical
spacetimes will scale in any straightforward way with increases in the
central mass.)  There are several ways one might go about trying to
track the region as the construction progresses.  One obvious, simple
way is by the triangulation of distances from some ``fixed'' markers
in the laboratory.  Because the metrical structure within the lab is
constantly changing, however, and doing so in very complex ways during
the periods when new matter is being introduced and distributed, and
the concomitant metrical perturbations are radiating away, there is no
canonical way of implementing the triangulation procedures; in fact,
the different ways of doing so are exactly captured by the different
families of frames one can fix to identify points among the members of
the ancestral family of spacetimes (which in this case, recall, now
respectively represent the spacetime region enclosed by the laboratory
at different stages of the construction of the central body).
According to some of the concrete implementations of the triangulation
procedure, \emph{i}.\emph{e}., according to different families of
frames one uses to identify points among the several members of the
ancestral family, the patch one tries to track will end up inside the
central body; according to other procedures, it will end up outside
the central body.  In consequence, what one means by ``the set of
spacetime points composing a small region at a fixed spatiotemporal
position relative to the central body'' will depend sensitively on how
one fixes and tracks relative spatiotemporal positions, which is to
say, depends sensitively on one's knowledge of the spacetime's
metrical structure.\footnote{One might object that, in this example,
  the experimentalist is really trying to track ``the same points
  through space over time'', not ``the same spatiotemporal points in
  different spacetimes''.  In fact, though, since the goal of the
  investigation is to determine how global metrical structure in
  slightly aspherical spacetimes differ for different values of the
  central mass, it is natural for the experimentalist to consider each
  static phase of the laboratory---the period after the last bit of
  mass has been added and the perturbations have settled down, but
  before the next bit of mass is added---as a separate spacetime in
  its own right, for the purposes of comparison.  An appropriate
  analogue is the so-called ``physical process'' version of the First
  Law of black-hole mechanics
  (\citeNP{wald-gao-proc-1st-genl-2nd-charged-rot-bhs},
  \citeNP{wald-qft-cst}), where one must identify two separate
  spacetimes (in the sense of two different solutions to the Einstein
  field equation) that differ in that one conceives of the one as the
  result of a dynamical evolution of the other, even though there is
  no concrete representation of that evolution as occurring in a
  single spacetime.}

We are finally in a position to offer a precise criterion for
``existence of spacetime points independent of metrical structure''
natural to the investigative contexts we have considered.  There are
in fact two natural criteria that suggest themselves, one weaker than
the other.  The first, suggested by the example of complexification
and stated somewhat loosely, is
\begin{defn}
  \label{def:limiting-criterion-identify}
  Points in a spacetime manifold \emph{have existence independent of
    metrical structure} if there is a canonical method to identify
  spacetime points during gradual modifications to the local spacetime
  structure.
\end{defn} 
My discussion of the example of complexification shows that, in this
context and using this criterion, spacetime points do not have
existence independent of metrical structure.

Now, based on the discussion of simplification, I propose a second
criterion, stronger than the first and formulated more precisely and
rigorously.  Fix an envelopment of a limiting family with a definite
limit space.  I say that a point in $\mathcal{M}_1$ with an associated
degenerate family of frames \emph{vanishes} (or that the point itself
is a \emph{vanishing point}) with respect to the given family of
frames.  I say that a point in $\partial \hat{\mathfrak{M}}$
\emph{appears} if there is no family of frames that converges to it.
\begin{defn}
  \label{def:limiting-criterion}
  Points in a spacetime manifold \emph{have existence independent of
    metrical structure} if no specification of a family of frames in
  any ancestral family of the spacetime has vanishing or appearing
  points.
\end{defn} 
I do not demand that one be able to identify in a preferred way a
spacetime point in the limit with any point of any member of one of
its ancestral families, much less for all its ancestral families; this
allows us to hold on to diffeomorphic freedom in the presentation of
the limit space.  I do not even demand that the criterion hold for
every possible spacetime model---perhaps in some spacetimes it makes
sense to attribute existence to spacetime points independent of
metrical structure, whereas in others (say, completely homogeneous
spacetimes) it does not.  I demand only that, for a given spacetime,
one not be able to make points in any of its ancestral families vanish
and not be able to make points in it, as the limit space, appear.
This attempts to capture the idea that, when we construct a spacetime
model and treat it as an idealized representation of a more complex
system---as it always is---then we can reliably identify spacetime
points in our model with points in the more complex system, albeit up
to diffeomorphic presentation.  If we cannot do this irrespective of
the more complex model we start from, then we cannot without
arbitrariness and artifice regard results of an investigation in the
context of the idealized model as relevant to the physics of the more
complex system, for we will be unable to identify the regions in the
more complex system that the results of the idealizing investigation
pertain to.  The example of Schwarzschild spacetime as a limit of a
family of Reissner-Nordstr\"om spacetimes clearly does not satisfy the
criterion, for there are points that vanish in the limiting procedure
(\emph{e}.\emph{g}., those in the shaded region of
figure~\ref{fig:reiss-nord}).  One may suspect that the existence of
singular structure in the two spacetimes fouls things up.  The
following result, however, establishes that no spacetime satisfies the
criterion, \emph{i}.\emph{e}., that its failure is universal and
depends on no special properties of any spacetime model.

Every spacetime has at least one ancestral family, the trivial one
consisting of the continuous sequence of itself, so to speak.
Construct an envelopment $\mathfrak{M}$ for it, with it itself as the
limit space, and apply a slight twist, so to speak, to every metric in
every model in the family so as to render each model non-isometric to
any other, \emph{i}.\emph{e}., so as to render the family non-trivial.
(One can make this idea precise in any of a number of simple ways,
such as using a smoothly varying 1-parameter family of linear
perturbations.)  On a curve in $\mathfrak{M}$, fix a family of frames
that has a well defined limit on $\partial \hat{\mathfrak{M}}$.  Now,
define a family of Lorentz transformations along that curve, one
transformation at each point, such that the family varies smoothly
along the curve, and such that when one applies each transformation to
the tetrad at its point, the result is a family of frames that has no
well defined limit.  (One can always do this; for example, the Lorentz
transformations can cause the tetrads to oscillate wildly as $\lambda
\rightarrow 0$.)  The points of the ancestral family along that curve
have no corresponding point in the limit space defined by the
resulting family of frames.  This proves
\begin{prop}
  \label{prop:vanish}
  Every spacetime has a non-trivial ancestral family with vanishing
  points.  Every non-trivial ancestral family has a limit space with
  respect to which some of its points vanish.
\end{prop}
In consequence, in every relativistic spacetime we treat as an
idealized model in the context of this sort of scientific
investigation, we can attribute existence to individual spacetime
points (or not), only by reference to the metrical structure of the
ancestral family we use to construct the model, and the limiting
process we choose for the construction.

An obvious objection to the relevance of these arguments to the ontic
status of spacetime points is that I deal here only with idealizations
and approximations, not with ``a real model of real spacetime''.  But
we never work with anything that is not an idealization---it's
idealizations all the way down, young man, as part of the human
condition.  If you can't show me how to argue for the existence of
spacetime points independently of metrical structure using our best
scientific theories \emph{as they are actually used in successful
  practice}, then you are not relying on real science to ground your
arguments.  You are paying only lip-service to the idea that science
should ground these sorts of metaphysical issues.

\section{Pointless Constructions}
\label{sec:pointless}

\resetsec

The argument of \S\ref{sec:limits} yields a conclusion that holds only
in a limited sphere, \emph{viz}., those investigations based on the
idealization of models of spacetime by means of limits.  One may
wonder whether it could be parlayed into a more general argument.  I
do not think so.  Indeed, I think there is \emph{no} sound argument to
the effect that no matter the context of the investigation one can
identify spacetime points or attribute existence to them only by
reference to prior metrical structure.  Sometimes, in some contexts,
one can attribute existence to them and identify them without any such
reference.  To show this, I will present an argument that all the
structure accruing to a spacetime, considered simply as a differential
manifold that represents the collection of all possible (or, depending
on one's modal predilections, actual) physical events, can be given
definition with clear physical content in the absence of metrical
structure.  The argument takes the form of the construction of the
point-manifold of a spacetime, its topology, its differential
structure and all tensor bundles over it from a collection of
primitive objects that, when the construction is complete, acquires a
natural interpretation as a family of covering charts from the
manifold's atlas, along with the families of bounded, continuous
scalar fields on the domain of each chart.  That idea yields the
following precise criterion the argument will rely on.
\begin{defn}
  \label{def:pointless-criterion}
  Points in a spacetime manifold \emph{have existence independent of
    metrical structure} if the manifold can be constructed from a
  family of scalar fields, the values of which can be empirically
  determined without knowledge of metrical structure.
\end{defn}

The basic idea of the construction is simple.  I posit a class of sets
of rational numbers to represent the possible values of physical
fields, with a bit of additional structure in the form of primitive
relations among them just strong enough to ground the definition of a
derived relation whose natural interpretation is ``lives at the same
point of spacetime as''.  A point of spacetime, then, consists of an
equivalence class of the derived relation.  The derived relation,
moreover, provides just enough rope to allow for the definition of a
topology and a differential structure on the family of all equivalence
classes, and from this the definition of all tensor bundles over the
resultant manifold, completing the construction.  The posited
primitive and derived relations have a straightforward physical
interpretation, as the designators of instances of a schematic
representation of a fundamental type of procedure the experimental
physicist performs on physical fields when he attempts to ascertain
relations of physical proximity and superposition among their observed
values.  An important example of such an experimental procedure is his
use of the observed values of physical quantities associated with
experimental apparatus to determine the values of quantities
associated with other systems, those he investigates by use of the
apparatus.  This interpretation of the relations motivates the claim
that the constructed structure suffices, for our purposes, as a
representation of spacetime in the context of a particular type of
experimental investigation as modeled by mathematical physics, and is
not (only) an abstract mathematical toy.

I begin the construction by laying down some definitions.  Let
$\mathbb{Q}$ be the set of rational numbers.  A \emph{simple pointless
  field} $\phi$ (or just \emph{simple field}) is a disjoint union
$\displaystyle \biguplus_{p \in \mathbb{Q}^4} f_p$, indexed by the set
$\mathbb{Q}^4$, such that
\begin{enumerate}
    \item\label{item:field-rational_vals}every $f_p \in \mathbb{Q}$
    \item\label{item:field-all-q4}there is an $f_p \in \phi$ for every
  $p \in \mathbb{Q}^4$
    \item\label{item:field-bounded}there are two strictly positive
  numbers $B_\text{l}$ and $B_\text{u}$ such that $B_\text{l} < |f_p|
  < B_\text{u}$ for all $p \in \mathbb{Q}^4$
    \item\label{item:field-continuous}the function $\bar{\phi}:
  \mathbb{Q}^4 \rightarrow \mathbb{Q}$ defined by $\bar{\phi}(p) =
  f_p$ is continuous in the natural topologies on those spaces, except
  perhaps across a finite number of compact three-dimensional
  boundaries in $\mathbb{Q}^4$
\end{enumerate}
Our eventual interpretation of such a thing as a candidate result for
an experimentalist's determination of the values for a physical field
motivates the set of conditions.  That we index $\phi$ over
$\mathbb{Q}^4$ means that we assume from the start that the
experimentalist by the use of actual measurements and observations
alone can impose on spacetime at most the structure of a countable
lattice indexed by quadruplets of rational numbers (and even this only
in a highly idealized sense); in other words, the spatiotemporal
precision of measurements is limited.
Condition~\ref{item:field-rational_vals} says that all measurements
have only a finite precision in the determination of the field's
value.  Condition~\ref{item:field-all-q4} says that the field the
experimentalist measures has a definite value at every point of
spacetime.  Condition~\ref{item:field-bounded} says that there is an
upper and a lower limit to the magnitude of values the experimentalist
can attribute to the field using the proposed experimental apparatus
and technique; for instance, any device for the measurement of the
energy of a system has only a finite precision, and thus can attribute
only absolute values greater than a certain magnitude, and the device
will be unable to cope with energies above a given magnitude.
Condition~\ref{item:field-continuous} tries to capture the ideas that
(local) experiments involve only a finite number of bounded physical
systems (apparatuses and objects of study), and that classical
physical systems bear physical quantities the magnitudes of which vary
continously (if not more smoothly), except perhaps across the
boundaries of the systems.

Fix a family $\Phi$ of simple pointless fields.  The \emph{link at
  $p$}, $\lambda_p$, is a set containing exactly one element from each
simple field in $\Phi$ such that all the elements are indexed by $p$,
the same quadruplet of rational numbers.  One link, for example,
consists of the set of all values in the fields in $\Phi$ indexed by
$(3/17, \, 2, \, -3001\frac{90}{91}, \, 2)$.  A \emph{linked family of
  simple pointless fields} $\mathfrak{F}$ is an ordered pair $(\Phi,
\, \Lambda)$ where $\Phi$ is a countable collection of simple fields,
and $\Lambda$ is the family of links on $\Phi$, a \emph{linkage},
complete in the sense that it contains exactly one link for each $p
\in \mathbb{Q}^4$.  The idea is that the values of the simple fields
in the same link all live ``at the same point of spacetime'', namely
that designated by $p$.  One can think of the linkage as a coordinate
system on an underlying, abstract point set.

We are almost ready to define the point-structure of the spacetime
manifold.  We require only two more constructions, which I give in an
abbreviated fashion so as to convey the main points without getting
bogged down in unnecessary technical detail.  Let $\mathfrak{F} =
(\Phi, \, \Lambda)$ be a linked family containing all simple fields;
we call it a \emph{simple fundamental family}.  Let
$\hat{\mathfrak{F}} = (\hat{\Phi}, \, \hat{\Lambda})$ be another.  We
want a way to relate the linkages of the two, so as to be able to
represent the relation between the coordinate systems of two different
charts on the same neighborhood of the spacetime manifold, or on the
intersection of two neighborhoods.  A \emph{cross-linkage} on a simple
fundamental family is an ordered triplet $(O, \, \hat{O}, \, \chi)$
where $O \subseteq \mathbb{Q}^4$ and $\hat{O} \subseteq \mathbb{Q}^4$
are open sets, such that either both are the null set or else both are
homeomorphic to $\mathbb{Q}^4$, and $\chi$ is a homeomorphism of $O$
to $\hat{O}$.  The link $\lambda_p \in \Lambda$ for $p \in O$, then,
will designate the same point in the underlying manifold as
$\hat{\lambda}_{\chi(p)} \in \hat{\Lambda}$ for $\chi(p) \in \hat{O}$;
in this case, we say the links \emph{touch}.  If $O$ and $\hat{O}$ are
the null set, then the represented neighborhoods do not intersect.
(We do not require that the values of the scalar fields in the two
different simple fundamental families be numerically equal at any
given point, as the two scalar fields may represent different physical
quantities, \emph{e}.\emph{g}., a component of the fluid velocity and
a component of the shear-stress tensor of a viscous fluid.)  One can
extend the idea of a cross-linkage to an arbitrary number of simple
fundamental families in the obvious way.  (To make the idea precise we
would need to index the collection of families, and so on, but I think
it is clear enough without going through the bother.)  We would then
identify a point in an underlying abstract point-set as an equivalence
class of links under the equivalence relation ``touches''.

To finish the preparatory work, we must move from rationals to reals.
Fix a simple, fundamental family $\mathfrak{F}$.  First, we attribute
to $\mathfrak{F}$ the algebraic structure of a module over
$\mathbb{Q}$.  For example, the sum of two simple pointless fields
$\phi$ and $\psi$ in $\Phi$ is a simple pointless field $\xi$ such
that $x_p \equiv f_p + g_p$ is the value in $\xi$ labeled by the index
$p$, where $f_p \in \phi$ and $g_p \in \psi$.  $\xi$ is clearly itself
a simple pointless field with a natural embedding in the linkage on
$\mathfrak{F}$, and so belongs to $\Phi$.  Now, roughly speaking, we
take a double Cauchy-like completion of $\Phi$ over both the points $p
\in \mathbb{Q}^4$ and the values $f_{\hat{p}} \in \mathbb{Q}$,
yielding the family $\bar{\Phi}$ of all disjoint unions of real
numbers continuously indexed by quadruplets of real
numbers.\footnote{In order to get the completion we require, standard
  Cauchy convergence does not in fact suffice.  We must rather use a
  more general method, such as Moore-Smith convergence based on
  topological nets.  The technical details are not important.  See,
  \emph{e}.\emph{g}., \citeN[ch.~2]{kelley-gen-topo} for details.}
This procedure makes sense, because every continuous real scalar field
on $\mathbb{R}^4$ is, again roughly speaking, the limit of some
sequence of bounded, continuous rational fields defined on
$\mathbb{Q}^4$.  We thus obtain what is in effect the family
$\bar{\Phi}$ of all continuous real scalar fields on $\mathbb{R}^4$,
though I refer to them as \emph{pointless fields}, in so far as, at
this point, they are still only indexed disjoint unions.  The limiting
procedure, moreover, induces on $\bar{\Phi}$ the structure of a module
over $\mathbb{R}$ from that on $\Phi$.  Finally, in the obvious way,
we take the completion, as it were, of $\Lambda$ using the same
limiting procedure to obtain a linkage $\bar{\Lambda}$ on
$\bar{\Phi}$.  I call $\bar{\mathfrak{F}} = (\bar{\Phi}, \,
\bar{\Lambda})$ a \emph{fundamental family}.  A cross-linkage on a
pair of fundamental families is the same as for simple fundamental
families, except only that one uses homeomorphisms on subsets of
$\mathbb{R}$ rather than $\mathbb{Q}$.  If we have two simple
fundamental families with a cross-linkage on them and take limits to
yield two fundamental families, then the nature of the limiting
process guarantees a unique cross-linkage on the two fundamental
families consistent with the original.

We can at last construct a real topological manifold from a collection
of simple fundamental families.  The basic idea is that a fundamental
family represents the family of continuous real functions on the
interior of a bounded, normal neighborhood of what will be the
spacetime manifold.  Because a spacetime manifold must be paracompact
(otherwise it could not bear a Lorentz metric), there is always a
countable collection of such bounded, normal neighborhoods that cover
it.  This suggests
\begin{defn}
  \label{def:pointless-mnfld}
  A \emph{pointless topological manifold} is an ordered pair $(\{
  \mathfrak{F}_i \}_{i \in \mathbb{N}}, \, \chi)$ consisting of a
  countable set of simple fundamental families and a cross-linkage on
  them.
\end{defn}
To justify the definition, I sketch the construction of the full
point-manifold and its topology.  First, we take the joint limit of
all simple fundamental families to yield a countable collection of
fundamental families with the induced cross-linkage.  A point in the
manifold, then, is an equivalence class of links, at most one link
from each family, under the equivalence relation ``touches''.  The set
of links associated with one of the families, then, becomes a
representation, with respect to the equivalence relation, of the
interior of a compact, normal neighborhood in the manifold, and the
fields in that family represent the collection of continous real
functions on that neighborhood.  The cross-linkage defines the
intersections among all these neighborhoods, yielding the entire
point-set of the manifold.  By assumption, the collection of all such
neighborhoods forms a sub-basis for the topology of the manifold, and
so, by constructing the unique topological basis from the given
sub-basis, the point-set becomes a true topological manifold.  It is
straightforward to verify, for example, that a real scalar field on
the constructed manifold is continuous if and only if its restriction
to any of the basic neighborhoods defines a field in the fundamental
family associated with that neighborhood.  

Now, to complete the construction, we can define the manifold's
differential structure in a straightforward way using similar
techniques.  First, demarcate the family of smooth scalar fields as a
sub-set of the continuous ones, which one can do in any of a number
straightforward ways with clear physical content based on the idea of
directional derivatives.  The family of all smooth scalar fields on a
topological manifold, however, fixes its differential structure
\cite{chevalley46}.  The directional derivatives themselves suffice
for the definition of the tangent bundle over the manifold, and from
that one obtains all tensor bundles.

After so much abstruse and, worse, tedious technical material, we can
now judge whether the construction supports the argument I want to
found on it.  The use of $\mathbb{Q}^4$ to index a simple pointless
field represents the fact that all points in a laboratory have been
uniquely labeled by 4 rational numbers, say, by the use of rulers and
stop-watches.  Such an operation neither measures nor relies on
knowledge of metrical structure, for it yields in effect only a chart
on that spacetime region.  (No assumption need be made about the
``metrical goodness'' of the rulers and clocks.)  Neither does any
other operation used in the construction pertain to metrical
structure.  One determines the values of the simple fields, for
example, by use of physical observations, which do not themselves
necessarily depend on knowledge of the ambient metrical structure.  To
illustrate the idea, consider the use of a gravity gradiometer to
measure the components of the Riemann tensor in a region of spacetime,
which exemplifies many of the ideas in the construction.  The
gradiometer is essentially a sophisticated torsion balance for
measuring the quadrupole (and higher) moments of an acceleration
field.\footnote{See, \emph{e}.\emph{g}., \citeNP[\S16.5,
  pp.~401--402]{misner-et73}, for a description of the device and its
  use.}  Its fixed center and the ends of its two rotatable axes
continuously occupy at any given moment 5 proximate points, the
attribution to which of values for linear and angular acceleration
yields direct measures of the components of the Riemann tensor in a
normal frame adapted to the position and motion of the instrument.
One then identifies the spacetime points the parts of the instrument
respectively occupy, and by extension those in the normal frame
adapted to it, by the values of the components of the Riemann tensor
and their derivatives in that frame, by the values of its scalar
invariants, and so on.\footnote{See, for example, Bergmann and
  Komar~\citeyear{komar-bergmann-pois-brck-loc-obs-gr,bergmann-komar-obs-commut}
  for a concrete, albeit purely formal, example of a procedure for
  implementing this idea.}  One does not have to postulate a prior
metric structure in order to perform the measurements and label the
points, nor need one have already determined the metrical structure by
experiment.  Indeed, in the performance of the gradiometer
measurements one determines much of spacetime's metrical structure.
Because the facts of intrinsic physical significance that the values
of the fields and the relations among them embody (is this body in
contact with another?\ does heat flow from that body to this or
vice-versa?), moreover, remain invariant under the action of a
diffeomorphism it follows that the equivalence classes we used to
construct points does so as well.  Thus, we can fix all the manifold
structure, including metrical, only up to diffeomorphism, as we
expect.  This shows that the construction delivers everything we need
and nothing more.

There is an obvious response to the argument based on this
construction.  One may object that, so far from the argument's having
shown that the construction pushes us to attribute independent
existence to spacetime points, it rather suggests that points are
defined only by reference to prior physical systems, and hence exist
in only a Pickwickian sense, dependent on the identifiability of those
physical systems.  This objection can be answered by, as it were,
throwing away the ladder.  Once one has the identification of
spacetime points with equivalence classes of values of scalar fields,
one can as easily say that the points are the objects with primitive
ontological significance, and the physical systems are defined by the
values of fields at those points, those values being attributes of
their associated points only \emph{per
  accidens}.\footnote{\citeN{stachel-mean-gen-covar-hole} provides an
  elegant tool for describing the result of such a construction as I
  propose and in particular this rebuttal to the proposed objection
  (though I should say his work is not related to a project such as
  this).  In his terms, I have sketched the construction of an
  \emph{individuating field} independent of the stipulation of
  metrical structure, \emph{viz}., a field or system of fields on
  spacetime that suffices for the identification of individual
  spacetime points.}  I do not pretend to endorse such a move, but I
do not have to.  My constructive argument is \emph{ad hominem}.

\section{The Debate between Substantivalists and Relationalists}
\label{sec:subs-rel}

\resetsec

I do not consider the idea of pointless manifolds deep or of great
interest in its own right.\footnote{There are a few questions of
  potential interest that accrue to it.  Is it possible to determine
  the topology of a non-compact manifold by the postulation of a
  finite number of simple fields?  If so, does the minimum number
  depend on a topological invariant?  Is it in any case greater than
  the number of fields we currently believe to have physical import?}
There are, I am sure, many other constructions in the same spirit.  If
one were so inclined, I suppose one could try to take something like
it to give a precise way for a relationalist to characterize the
spacetime manifold.\footnote{See \citeN{butterfield-relat-poss-wrlds}
  for a survey of some ways one might attempt such a project.}  I am
not so inclined, because I do not think the contemporary debate
between the relationalist and the substantivalist has been well posed,
and I am inclined to think it never will be in any interesting sense.
That is what I take to be the force of the opposed constructions of
\S\ref{sec:limits} and \S\ref{sec:pointless}, taken in tandem.  They
show that ``dependence on prior metrical structure'' is formal,
\emph{i}.\emph{e}., without substantive content until given
explication in the framework of an investigative enterprise, even if
that framework be given only in schematic form.  Once one grants this,
however, the game is up.  Different investigative frameworks can and
do yield natural criteria that lead to contrary
conclusions.\footnote{This line of argument bears fruitful comparison
  to the ideas of \citeN{ruetsche-interp-quantth}, though it was
  developed independently of her work.}

An amusingly poignant feature of the constructions shows this clearly:
each yields a conclusion contrary to what the traditional debates
would have led one to have expected based on the tools and techniques
it employs.  In the second, one uses independent values of physical
quantities (a stock in trade of the relationalist) in order to
identify and attribute existence to spacetime points without a prior
assumption of metric structure; and in the first, one uses structures
in mathematical physics that seem to presuppose the independent
identifiability of spacetime points (a stock in trade of the
substantivalist) in order to argue that in fact they are not
identifiable without a prior postulation of metric structure.  One may
think that these features of the arguments make them, in the end,
self-defeating, but I do not think that is so.  In the first, one
operates under the implicit assumption that the more complex models
one idealizes are themselves only idealizations of yet more complex
models.  In the second, one implicitly assumes that, say, the
gradiometer is small enough and the temporal interval of the
measurement itself short enough to justify the use of the Minkowski
metric in making the initial attributions of the magnitudes of
spatiotemporal intervals in the experiment; one then uses this to
bootstrap one's way to a more accurate representation of the metrical
structure of spacetime, which is what is done in practice.  I think
that this facet of the arguments, perhaps more than anything else,
illustrates the vanity of the traditional debate: one can use the
characteristic resources and moves of each side to construct arguments
contrary to it, once one takes the trouble to make the question
precise.

Most damning in my eyes, the constructions show the futility of the
debate, for they make explicit how very little one gains in
comprehension or understanding by having taken the considerable
trouble to have made the questions precise.  Indeed, one may feel with
justice that nothing has been gained, but rather something has been
lost in a pettifoggery of irrelevant technical detail.\footnote{Jeremy
  Butterfield in particular has vigorously tried to convince me that I
  dismiss too readily the possible philosophical value of the
  technical constructions and arguments of \S\S\ref{sec:limits} and
  \ref{sec:pointless}.  I would like to think he is right.}

Although I conclude the traditional debate is without real content, I
think there is a related, interesting question one \emph{can} give
clear sense to: what in one's investigative framework is naturally
taken to, or must one take to, have intrinsic physical significance?
Even putting aside existence and ontology as emotive distractions,
however, I do not think one can give even this question substantive
sense in the abstract: the question is a formal template that one must
give substance to by fixing the significance of its terms in
presumably different ways in different particular contexts.

Consider one way to rephrase the question that may seem on its face to
give it concrete content in abstraction from any schematic framework:
what propositions would all observers agree on?  One cannot answer
this question in the abstract, or even give it definite sense, because
one has not yet fixed the way that one will schematically represent
the observer (or experimental apparatus) and the process of
observation.  In order to do so, one must settle many questions of a
more concrete nature.  Will one use the same theory to model the
observation as one uses to model the system?  Will one take the
observer to be a test system, in the sense that the values of its
associated physical quantities do not contribute to the initial-value
formulation of the equations of motion of one's theoretical or
experimental models?  And so on.  Until one settles such issues, one
cannot even say with precision what any single observer can or will
observe, much more what all will agree on.  In this sense, even claims
such as ``in general relativity, only what is invariant under
diffeomorphisms has intrinsic physical significance'' have only
schematic content.  One must give definite substance to the ``what''
in ``what is invariant''---substance that involves the forms of the
physical systems at issue and the methods available for their probing
and representation---before one can make the claim play any definite
role in our attempts to comprehend the world.  I take this to be the
lesson of \citeN{stein-phil-prehist-gr}, \emph{viz}., that the way to
proceed in these matters is the one Newton and Riemann relied on: we
must infer what we can about the spatiotemporal structure of the world
from the roles it plays in characterizing physical interactions; and
on this basis, neither substantivalism nor relationalism can claim any
great victory.\footnote{\citeN[p.~274]{disalle-dyn-indiscern-st}
  trenchantly makes a very closely related point, one, indeed, that in
  large part may be viewed as foundational for my analysis:
  \begin{quote}
    Since the work of Riemann and Helmholtz, however (not to mention
    Einstein), it should be clear that our claims about `objective'
    spatiotemporal relations always involve assumptions about the
    physical processes we use for measurement and stipulations about
    how those processes are to indicate aspects of geometry.
  \end{quote}}

In the end, why should we ever have expected there to have been a
single, canonical way to explicate the physical significance of the
idea of a spacetime point, on the basis of which we might then attempt
to determine whether such a thing exists or not in some lofty or
mundane sense?  What, after all, is lost to our comprehension of the
physical world without such a unique, canonical explication?  We
purport after all, in these debates, to attempt to better comprehend
the \emph{physical} world.  Hadn't we better ensure, then, that the
terms of our arguments have the capacity to come in some important way
into contact with the physical world by way of experiment and theory?
Once we take that demand seriously, we find an orgiastic crowd of
possible candidates to serve as concrete realizations of the question,
some of which will be fruitful in some kinds of enterprises, others in
others, and, most likely, several in none at all.  Indeed, I am far
from convinced that the question of the existence of spacetime points
has ever itself been well posed.  I think a necessary (though not
sufficient) condition for the scientific cogency and relevance of that
question is a demonstration that an answer to it would contribute
fruitfully to the proper comprehension of the performance of an
experiment or the proper construction of a model of a physical system
in the context of general relativity.  (Recall this paper's epigraph
by Maxwell.)  But what possible difference could an answer to it make
one way or another to those scientific issues?

I think there is a better question at hand: what mathematical
structures ``best'' represent our experience of spatiotemporal
localization?  Again, this question cannot be answered in the
abstract, for it depends sensitively on the answers to other, more or
less independent and yet inextricable questions, such as: what
mathematical structures best represent our experience of other
features of spatiotemporal phenomena, such as the lack of absolute
simultaneity, the orientability of space, \emph{etc}.?  And also
questions such as: what structures for representation of various kinds
of derivatives do we need to formulate equations of motion?  And what
structures for representation of Maxwell fields?  And so on.  One has
to attempt to address these questions in a dialectical fashion,
answering part of one here, seeing what adjustments that requires in
other parts of the manifold of possible structures, so to speak, and
so on.  The answer to one of these questions in one context may be
individual points of a spacetime manifold, to another question in
another context it may be area and volume operators as in loop quantum
gravity, and so on.  Instead of asking whether the manifold itself or
the manifold plus the metric is ``\emph{really} spacetime'', we should
rather be asking what sorts of structure with real physical
significance a manifold by itself and a manifold with a metric can
respectively support---anything requiring only differential topology
or geometry for the former, and anything requiring Lorentzian geometry
for the latter.  It is to the investigation of such questions that I
now turn.

\section{An Embarassment of Spacetime Structures}
\label{sec:embarass}
\resetsec

The arguments of this paper extend themselves naturally beyond the
realm of the debate over the existence of spacetime points, and do so
in a way that sheds further light on the futility of that debate.
There are many different senses one can give to the question whether
some putative entity or structure of any type has real physical
significance in the context of general relativity, each more or less
natural in different contexts.  For lack of a better term, I shall say
that an entity (which, as we shall see, can encompass several
different types of thing), purportedly represented by a theoretical
structure, has \emph{physicality} if one has a reason to take that
structure seriously in a physical sense, \emph{viz}., if one can show
that it plays an ineliminable, or at least fruitful and important,
role in the way that theory makes contact with experiment.  Of course,
as I stressed in \S\ref{sec:hole}, such an abstract, purely formal
schema as ``plays an ineliminable, or at least fruitful and important,
role in the way that theory makes contact with experiment'' has no
real content until one explicates it in the context of a more or less
well delineated investigative framework.  It is, in fact, one of the
``important matters on which sensible even if vague things can be
said,'' which Stein discussed in the passage I quoted on
page~\pageref{pg:stein-struc-know}.  As such, it is the examples that
give the idea life.

\subsection{Manifest Physicality}
\label{sec:manifest-physicality}

A Maxwell field, represented by the Faraday tensor $F_{ab}$, is
manifestly physical.  One important sense in which this is true turns
on the fact that it contributes to the stress-energy tensor on the
righthand side of the Einstein field equation.  The Maxwell field
itself possesses stress-energy, and in general relativity nothing is
physical if not that.

Consider now a Killing field on spacetime, a vector field $\xi^a$ that
satisfies Killing's equation
\begin{equation}
  \label{eq:kill-eqn}
  \nabla_{(a} \smidge \xi_{b)} = 0
\end{equation}
and so generates an isometry, in the sense that $\pounds_\xi \smidge
g_{ab} = 0$.  In this guise, it seems not to possess the
characteristics of a physical field, in so far as it enters the
equations of motion of no manifestly physical system, such as a
Maxwell field.  In other words, it does not couple with phenomena we
consider physical, does not contribute to the stress-energy tensor.
Now, define the 2-index covariant tensor $P_{ab} \equiv \nabla_a
\smidge \xi_b$.  Equation~\eqref{eq:kill-eqn} implies that it is
anti-symmetric.  Let us say that it happens as well to have vanishing
divergence and curl, $\nabla_n P^{na} = 0$ and $\nabla_{[a} P_{bc]} =
0$, and so satisfies the source-free Maxwell equations.  Is it
\emph{eo ipso} a true Maxwell field, and so physical?  Not
necessarily.  There are always an innumerable number of 2-forms on a
spacetime that satisfy the source-free Maxwell equations.  At most,
one of them represents a physical Maxwell field.  If, however, it just
so happened that $P_{ab}$ were to represent the physical Maxwell field
on spacetime---one known as a Papapetrou field in this case---the fact
that one natural way to represent the field happened to generate an
isometry would appear to be an accident, in the sense that no property
of the field accruing to it by dint of its physicality, which is to
say, by dint of its satisfaction of the Maxwell equations and
concomitant coupling with other manifestly physical phenomena (such as
spacetime curvature, by way of the Einstein field equation), depends
on the satisfaction of equation~\eqref{eq:kill-eqn} by $\xi^a$ (except
in the trivial sense that satisfaction of equation~\eqref{eq:kill-eqn}
is necessary for $\xi^a$ to be a 4-vector potential for a Maxwell
field).  Still, $\xi^a$ is a naturally distinguished geometrical
structure in the physical description of spacetime, forms a part of
the description of spacetime independent of the particulars of the
physical constitution of any observed phenomena, in particular in so
far as it places non-trivial contraints on a manifestly physical
structure, the spacetime metric.  In this sense, different from that
pertaining to the Maxwell field, $\xi^a$ is physical, for the Maxwell
field, by contrast, is not naturally distinguished in this sense, but
rather depends in an essential way on the peculiar, contingent
physical constitution of a particular family of phenomena.

In what sense, though, is the metric manifestly physical?  The metric
does not itself contribute to the stress-energy content of spacetime,
for one cannot attribute a localized gravitational stress-energy to
it.\footnote{See, \emph{e}.\emph{g}.,
  \citeN{curiel-geom-objs-nonexist-sab-uniq-efe}.}  That is not to say
that the metric does not appear in the stress-energy tensor of a given
spacetime, for it is almost always required for the construction of
the stress-energy tensor.\footnote{Indeed, the only example I know of
  a stress-energy tensor for which the metric is not needed for its
  definition is the case of a null gas, for which only the conformal
  structure of spacetime is required.  See
  \citeN{lehmkuhl-mass-energy-mom-st} for discussion of these issues.}
The stress-energy tensor of a Maxwell field, for example, is $F_{an}
F^n {}_b + \afourth g_{ab} F_{rs} F^{rs}$.  (The metric appears not
only explicitly in the second term, but also implicitly in both terms,
raising the contracted indices.)  The metric, however, is necessary
both for posing the initial-value formulation of every possible kind
of field that may appear in a relativistic spacetime, in particular
all of those (such as the Maxwell field) that we regard as manifestly
physical, and for formulating the equations of motion of the fields.
In particular, the metric dynamically couples with other physical
systems, \emph{i}.\emph{e}., enters into interaction with them in the
strong sense that there always exist terms in the equations of motion
for any given field in which the metric appears as one factor and the
tensor representation of the field as another.  For the Maxwell field,
the metric appears contravected with the Faraday tensor in the field
equation representing the fact that its covariant divergence equals
the charge-current density of matter.\footnote{That the other defining
  equation for a Maxwell field, representing the fact that the Faraday
  tensor is curl-free, does not require the metric at all for its
  formulation---the exterior derivative is determined by the
  differential structure of the underlying manifold, and does not
  require any other structure at all for its definition---may push one
  to say that it is not a dynamical equation of motion, but rather a
  kinematical constraint.}

The metric, of course, can play other roles as well, just as a Killing
field.  A vacuum spacetime with non-zero cosmological constant has a
stress-energy tensor equal to the metric times a constant.  In this
case, one plausible way of reading the Einstein field equation is to
have the metric play simultaneously two distinct roles, one as the
necessary ground of all spatiotemporal structure (embodied in the
Einstein tensor) and the other as a component in the tensor
representing the stress-energy content of spacetime, depending on
contingent features of the ambient matter field, in this case,
whatever field gives rise to the cosmological constant.  Again, in the
former sense, as ground of spatiotemporal structure, the metric is a
naturally distinguished structure in any physical description of
spacetime; in the latter sense, it rather depends on the peculiar,
contingent physical constitution of a particular family of phenomena.

Consider the Riemann tensor.  Again, it manifests physicality in
several different ways, in different contexts.  Perhaps the most
important is in the equation of geodesic deviation, where it directly
measures the rate at which infinitesimally neighboring geodesics tend
to converge towards or diverge away from each other.  In this case,
the Riemann tensor's physicality consists in the fact that it encodes
all information needed to model manifestly observable phenomena,
\emph{viz}., the relative acceleration of nearby freely falling
particles and the tidal force exerted between different parts of a
freely falling extended body.  Another important role it plays in
general relativity is as the measure of the failure of the ambient
covariant derivative operator associated with the spacetime metric to
commute with itself when acting on vectors or tensors.  Here, the
physical interpretation is not clear, but one way of trying to
explicate it is by considering the way that a tangent vector changes
when parallel-propagated around an ``infinitesimally small''
loop.\footnote{See \citeN[ch.~2, \S3]{wald-gr} for a thorough
  exposition.}  The infinitesimal change in the vector when it returns
to the initial point is directly proportional to the Riemann tensor.
Still, it is difficult to say that this has real \emph{physical}
significance, in so far as one could implement such a mechanism and
measure the result only in a spacetime with closed causal curves.  And
yet so much of the mathematical apparatus of general relativity
depends on the fact that the ambient derivative operator is, in
general, not flat (\emph{i}.\emph{e}., fails to commute with itself),
that it would be absurd to say that the Riemann tensor is not playing
a physical role here.  What exactly that role is, however, is not easy
to pin down.  This is an example of the kind of philosophically
important problem whose resolution would have manifest physical
significance that I take Maxwell to be referring to in the passage I
used as one of this paper's epigraphs.

The Einstein tensor itself presents an interesting case.  It has no
straightforward geometrical interpretation.\footnote{See
  \citeN[\S2.1]{curiel-primer-econds} for a discussion.}  It seems,
moreover, to have no straightforward physical interpretation
either---it enters into the equations of motion of no known fields; it
measures no quantitative feature of any known physical phenomena; it
does not represent a field possessing stress-energy; it constrains the
behavior of no other manifestly physical structure; and so on.  And
yet it is the structure that matter fields couple to (via the Einstein
field equation) in their role as source for spatiotemporal curvature.
In this role, it dynamically couples with no individual matter fields,
but rather only to the aggregate physical quantity ``stress-energy''
they all possess, and which, according to the fundamental principle of
the fungibility of all forms of energy,\footnote{See
  \citeN[ch.~\textsc{v}, \S97]{maxwell-matt-mot} and
  \citeN[chs.~\textsc{i, iii, iv, viii,
    xii}]{maxwell-theory-heat-1888} for illuminating discussion of
  this principle.}  in no way differs qualitatively among all known
fields.  Again, then, it seems manifestly physical in some sense, but
it is difficult to put one's finger clearly on that sense, and, again,
this is an example of a philosophically important problem whose
resolution would provide real physical insight.

Global structures of various sorts (causal, topological, projective,
conformal, affine, \emph{et al}.\@) present interesting cases as
well.\footnote{I take a structure to be global if it is not local in
  the sense explicated by \citeN[p.~55]{manchak-know-glob-struc-st}:
  \begin{quote}
    [A] condition $C$ on a spacetime is \emph{local} if, given any two
    locally isometric spacetimes $(M, \, g_{ab})$ and $(M', \,
    g'_{ab})$, $(M, \, g_{ab})$ satisfies $C$ if and only if $(M', \,
    g'_{ab})$ satisfies $C$.
  \end{quote}
  I think Manchak's definition of ``local'' is superior, as judged by
  its physical significance in the context of general relativity, to
  the one I proposed in \citeN[\S5]{curiel-sing}, though the latter
  may still be of interest in purely mathematical contexts, or in
  contexts of physical investigation that transcend the scope of a
  single theory.}  Consider the conformal structure of a spacetime.
It governs and is embodied in the relative behavior of the null cones
across all spacetime points.  One natural interpretation of the null
cones is as determining a finite, unachievable upper-limit for the
velocities of material systems.\footnote{See, however,
  \citeN{geroch-faster-light} and
  \citeN{earman-no-suplum-prop-class-q-flds} for dissenting
  arguments.}  The fact that the null cones determine a topological
boundary for the chronological future and past of every spacetime
point also has a natural interpretation in the same vein: if the
chronological future or past were topologically closed, then there
would be a limiting upper velocity for massive bodies that would be
\emph{actually achievable} by a massive body using only a finite
amount of energy.  If one accepts these interpretative glosses, then
the conformal structure has physicality in so far as it constrains the
behavior of manifestly physical systems.

So, to sum up, the notions of physicality mooted here are:
\begin{itemize}
    \item contributes to $T_{ab}$ (\emph{e}.\emph{g}., Maxwell field)
    \item required for initial-value formulation of manifestly
  physical fields (\emph{e}.\emph{g}., Maxwell field, $g_{ab}$)
    \item dynamically couples to manifestly physical entities
  (\emph{e}.\emph{g}., Maxwell field, $g_{ab}$)
    \item dynamically couples to manifestly physical quantities that
  more than one type of physical system can bear (\emph{e}.\emph{g}.,
  Einstein tensor)
    \item acts as a measure of an observable aspect of manifestly
  physical entities (\emph{e}.\emph{g}., Riemann tensor)
    \item enters the field equation of a manifestly physical structure
  (\emph{e}.\emph{g}., Einstein tensor)
    \item constrains the behavior of a manifestly physical entity
  (\emph{e}.\emph{g}., Killing field, conformal structure)
    \item plays an ineliminable, though physically obscure, role in
  the mathematical structure required to formulate the theory
  (\emph{e}.\emph{g}., Riemann tensor, Einstein tensor)
\end{itemize}
I am confident there are yet more senses of physicality I have not
touched upon.

\subsection{Observability}
\label{sec:observ}

One does not have to be an instrumentalist or an empiricist to accept
that the possible observability of physical phenomena is one of the
most fundamental reasons we have to think such things are physical in
the first place.  The question of the observability of various kinds
of global structure in general relativity, therefore, poses
particularly interesting problems for arguments about physicality.
Manchak~\citeyear{manchak-know-glob-struc-st,manchak-phys-reas-st}
shows that, in a precise sense, local observations can never suffice
to determine the complete global structure of spacetime in general,
and in particular cannot determine whether a spacetime is inextendible
or stably causal \cite[p.~418, proposition~3]{manchak-phys-reas-st}.
Nonetheless, there remain several things to say and ask about the
matter of physicality here.

Take, for example, the Euler number of the spacetime manifold, a
global topological structure.\footnote{See, \emph{e}.\emph{g}.,
  \citeN[ch.~\textsc{viii}]{alexandrov-comb-topol-2}.}  It is a
topological invariant that, in part, constrains the possible existence
of everywhere non-zero vector fields on a manifold.  That an
even-dimensional sphere, for example, possesses no everywhere non-zero
vector field (and indeed no Lorentzian metric) follows directly from
the computation of its Euler number.  If we were to live in a world
whose underlying manifold possessed a non-trivial Euler number, and so
could support no physical process that would manifest itself as an
everywhere non-zero vector field, this would constitute a physical
fact about the world in an indubitable sense.  It is not clear to me,
however, whether in some precise sense the Euler number of the
spacetime manifold could ever be determined by direct observation.

The orientability of spacetime is an example of a global topological
structure that seems to be strictly inobservable in an intuitive
sense.  This follows from the fact that one can construct an
orientable manifold from any non-orientable one by lifting the
structures on it to a suitable covering space, which is automatically
orientable.  The lift of the spacetime metric to a covering manifold,
however, would yield a representation of exactly the same physical
spacetime as the original: every physical phenomena in the one has an
isometric analogue, as it were, in the other, and vice-versa.  Whether
or not a spacetime manifold is simply connected, moreover, seems to be
in the same boat, for the universal covering manifold of a manifold is
guaranteed to be simply connected.\footnote{In order for a manifold to
  possess a universal covering manifold, it must be semi-locally
  simply connected.  Intuitively, this means that it cannot contain
  ``arbitrarily small holes''.  More precisely, it means that every
  point in the space has a neighborhood such that every loop in the
  neighborhood can be continuously contracted to a point.  (The
  contraction need not occur entirely with the given neighborhood.)
  The so-called Hawaiian Ear-Ring is an example of a topological space
  that is not semi-locally simply connected
  \cite{biss-genl-approach-fund-grp}.  Whether or not a spacetime
  manifold is semi-locally simply connected presents us with yet
  another type of question related to physicality: strictly speaking,
  there is no physical need for a manifold to possess a universal
  cover, and it is difficult, to say the least, to see what other
  physical relevance being semi-locally simply connected could have;
  and yet the construction of the universal cover is such an
  extraordinarily useful theoretical device \cite{geroch-topol-gr}
  that one wants to demand that a candidate spacetime manifold be
  semi-locally simply connected.  What status does such a demand have?
  A purely pragmatic one?}

Nonetheless, I think those answers about the possible observability of
a manifold's orientability and simple connectedness may be too pat.
If one were to observe that any member of a certain family of closed,
physically distinguished spatiotemporal loops could not be
continuously deformed into any member of another family of closed,
physically distinguished spatiotemporal loops, one would have shown
that the spacetime manifold is not simply connected.  Similarly, if
one could show that to parallely propagate a fixed orthonormal tetrad
around a given closed spatiotemporal loop would result in its
inversion, one would have demonstrated that spacetime is not
orientable.  I personally have no idea what sorts of experiment could
show either of those things.  The history of physics, however, if it
shows us nothing else, does show us never to underestimate the
ingenuity of experimentalists, no matter what the theoretician may
tell them is impossible to observe or measure.

The first Betti number of the spacetime manifold offers another
interesting example of this sort.  The first Betti number of a
topological space is the number of distinct connected components it
has; any manifold with a first Betti number greater than one is
\emph{ipso facto} not connected.  Say that we posited a non-connected
spacetime manifold.  According to the principles of general
relativity, any phenomena in one component would be strictly
inobservable in any other.\footnote{Perhaps one could posit some form
  of quantum entanglement among phenomena in the different components.
  The ramblings of many theorists of quantum gravity notwithstanding,
  such a possibility lies so far beyond the ambit of current well
  entrenched experimental technique and well founded theoretical
  knowledge as to render it incomprehensible as physics.  By the
  nature of the case, for instance, we could perform no direct
  experiments on the putatively entangled phenomena in the postulated
  other component to verify the entanglement beyond a shadow of a
  doubt.}  By this criterion, it makes no sense to attribute
physicality to regions of spacetime disconnected from our
own.

So, are these possibly inobservable global structures physical?  Well,
it seems to me that in one sense they are, and in others they are not.
The only lesson I want to draw here is that questions of this sort
require in-depth investigation sensitive both to the technical details
of the mathematics and to the physical details of how such structures
may and may not bear on other phenomena we think of as manifestly
physical, even if they turn out to be indubitably
inobservable.\footnote{The family of phenomena in relativistic
  spacetimes grouped under the rubric ``singular stucture'' (or
  ``singularities'') provides on its own a rich and diverse selection
  of examples, which I do not have room even to sketch here.  See
  \citeN{curiel-sing} for an extended discussion.}

\subsection{Physicality and Existence}
\label{sec:phys-exist}

What I have discussed so far in this section, I submit, are
philophically rich, scientifically significant questions and
arguments, of the sort Maxwell mentions in the epigraph to this paper.
Insight into and progress on any of the questions would constitute
real progress in our attempts to understand the world in a scientific
sense.  The sterility of the current debate between substantivalists
and relationalists is shown in the fact that no questions it addresses
has scientific value in the sense of Maxwell---it has spurred no work
with direct scientific, as opposed to purely metaphysical, import.

Still, No matter how convincing or interesting or philosophically rich
these examples and arguments may be, one might still want to respond
that they show nothing about the possible \emph{existence} of
spatiotemporal entities, and so in the end they do not bear on the
debate between substantivalism and relationalism.  I do not think that
is the correct lesson to leave with, though.  I take physicality to be
a necessary condition for the attribution of existence to a
theoretical entity.  If there are many possible ways an entity can
manifest physicality, therefore, and one can show that different
entities manifest some but not others of them, then it follows that it
is meaningless to attribute existence \emph{simpliciter} to such
theoretical entities.  If there are two entities each manifesting a
different type of physicality, then, in so far as each is a necessary
condition for existence, if one attributes existence to those
entities, it must be of a different sort for each.  Thus, in so far as
one wants to make sense of the idea of ``existence'' in the context of
physical entities purportedly represented by theoretical structures
(if that is the sort of thing one likes to do), it cannot be univocal.
To paraphrase Aristotle, existence is said, if at all in physics, in
many ways.

What light, if any, does all this shed on the cogency of the
traditional debate about the ontic status of spacetime?  I think quite
a bit.  A spacetime point is not physical in any of the ways I have
explicated: there is no such thing as an initial-value problem for
them; there is no equation of motion for them; no property of theirs
dynamically couples to any physical field; and so on.  How, then, is
one supposed to try to answer the question of whether or not they
exist in any way that purports to be grounded in physics?

\section{Valedictory Remarks on Realism and Instrumentalism, and the
  Structure of Our Knowledge of Physics}
\label{sec:valedict}

\resetsec

I think my conclusions about the vanity of metaphysical argumentation
abstracted from the pragmatics of the scientific enterprise carry over
into the general debate over realism and instrumentalism.  Indeed, I
consider the argument about relationalism and substantivalism to be an
instance of the more general form of argument one can give for
existence claims about entities and structures in science.  I will
consider two examples to make the point, the first somewhat
sophisticated, the second quite simple.

Consider, first, the Unruh effect.\footnote{See \citeN{wald-qft-cst}
  for a rigorous exposition of the phenomenon.}  The effect, roughly
speaking, is as follows.  (I discuss it only in the context of a
special case, but this does not affect the point.)  Consider two
observers in Minkowski spacetime pervaded by a scalar quantum field in
its vacuum state.  Each observer carries a simple particle-detector
coupled to the field, with two states: an excited state (``particle
observed''), and a ground state (``particle not observed'').  Both
detectors are initially in the ground state.  The first observer
follows a geodesic, and so does not accelerate; in this case, quantum
field theory predicts that the particle detector will remain in the
ground state, \emph{i}.\emph{e}., the probability that he will detect
any particles is zero, as one would expect on physical grounds, since
the background field is in the vacuum state.  The second observer,
however, begins to accelerate.  Now, there is a high probability that
her detector \emph{will} change from the ground to the excited state;
she will ``see particles''.  That is the Unruh effect.  Even though
the two observers disagree on whether there are particles or not, they
both agree that the state of the second particle detector changes, so
there is a physical fact of the matter in that sense.\footnote{Roughly
  speaking, the resolution of the paradox turns on the fact that an
  accelerating system in Minkowski spacetime occupies a negative
  energy-state: the accelerating detector, in dropping to an energy
  level beneath that of the ambient vacuum, registers the vacuum as
  having positive energy, which the accelerating observer interprets
  as having ``detected a particle''; the inertial observer, however,
  accounts for the drop in the accelerating detector's energy by
  concluding that it \emph{emitted} a particle, and so changed its
  state.  If one likes, one may take this as one way to make precise
  the idea that ``particle'' is not a natural notion in quantum field
  theory, and is indeed at times not only not useful but downright
  obfuscatory.}  Now, the bit of most interest to us is that the
fluctuations in the field that determine the change in the state of
the detector do not contribute to the definition of the stress-energy
tensor.  All observers, both inertial and accelerating, will still
conclude that the ambient stress-energy tensor is that of the vacuum
state.  Is Unruh radiance, then, physical or not?  Is it ``real''
radiation?  Well, in the sense that it is a phenomenon that all
observers will agree on, one that manifests itself in directly
observable effects, yes; in the sense that it does not contribute to
the stress-energy of spacetime, no.

Now, consider the question ``do electrons exist?''  On its face, it
seems immune to the sorts of problems I raise about the ontic status
of spatiotemporal structure.  Surely one can attribute canonical
significance to the question ``do electrons exist?'' independent of
investigative framework?  In fact, one cannot.  Think of the different
contexts in which the concept of an electron may come into play, and
the natural ways in those contexts one may want to attribute
physicality (or not) to electrons.  A small sample:
\begin{itemize}
    \item as a component in a quantum, non-relativistic model of the
  Hydrogen atom
    \item as an element in the relativistic computation of the Lamb
  shift
    \item as a possible ``constituent'' of Hawking radiation in an
  analysis of its spectrum
    \item as a measuring device in the observation of parton structure
  from deep inelastic scattering of electrons off protons, as modeled
  by the Standard Model
\end{itemize}
In the first case, one may want to attribute physicality to the
electron in so far as its associated quantities enter into the
initial-value formulation of the system's equations of motion; in the
second, one may base the attribution on the fact that one identifies
the electron as the bearer of definite values for the kinematic
Casimir invariants of spin and mass; there is no good definition in
general of an electron in the third, because there is no unambiguous,
physically significant definition of ``particle'' in quantum field
theory on a curved spacetime, and so \emph{a fortiori} no way to
attribute physicality to such a thing;\footnote{In essence, this is
  because one has no privileged group of timelike symmetries in a
  generic spacetime, as one has in Minkowski spacetime, on which to
  ground the notion of a particle.  See \citeN{wald-qft-cst} for a
  detailed explanation.}  in the fourth and final case, one can
attribute physicality to the electron because one can associate
localized charge, spin and lepton number with the mass-energy
resonance that represents it.  Now, one cannot even formulate in a
rigorous, precise way (and, indeed, often not even in a loose and
frowzy way) the criterion for physicality in any of these frameworks
in the terms of at least some of the others.

It follows that even in this case any formulation of the question in
abstract terms, such as ``what all observers agree on'' or ``what has
manifestly observable effects'' or ``what couples with other systems
we already think of as physical'' or ``what is essential to the
formulation of the theory'', remains empty until one renders content
to it by the fixation of a framework, even if only schematic.  To be
clear, I do not claim that one must always make the investigative
framework of one's work explicit, only that one ought to recognize it
must be there in the background, specifiable when push comes to shove,
as it will from time to time.  

In the picture I have implicitly relied on in the construction of my
arguments, the structure of physics may be thought of as something
like a differential manifold itself, with different techniques and
concepts that find appropriate application in different sorts of
investigation, and even in similar sorts of investigation of different
subject matters, all covering their own idiosyncratic patches of the
global manifold, consonant with each other when they overlap but with
none necessarily able to cover the entirety of the space.  In that
vein, I am confident there are many other interesting senses one can
render to the idea of the physicality of putative entities and
structures represented by our best physical theories, variously useful
or at least illuminating in investigations of different sorts.  In
some of those senses, one will rightly, or at least usefully or
suggestively, say those things are physical.  In others, one will not.
The words we use to further all the sorts of scientific and
philosophical investigations we pursue do not matter, only the
concepts behind the words, some of which find natural application in
some investigations and some of which do not.

This is not instrumentalism.  Among other things, I neither make nor
rely on any privileged claim about how one ought to understand the
structures of our best theories as formal systems, the terms and
relations with which we formulate them, and their broader or deeper
relation to the world itself, only about how we ought not understand
them.  The greatest physicists have always, it seems to me, had the
capacity to to think in both realist and instrumentalist ways about
both the best contemporary theories and the most promising lines of
theoretical attack as they were being developed.  Often, they held
both sorts of views in their minds at the same time, keeping many
avenues open, sometimes moving forward along one, sometimes switching
to another, sometimes straddling the line, as best befit the demands
of the investigation, with a concomitant gain in richness of
conception and depth of
thought.\footnote{Stein~\citeyear{stein-phil-prehist-gr,stein-locke-huy-newt,stein-poincare}
  forcefully argues this line of thought.}  In some contexts and for
some purposes it is most useful to conceive, think and speak in
realist terms, and in others to do so in instrumentalist terms.  They
are both good in their place, and neither is correct \emph{sub specie
  {\ae}ternitatis}.  In any event, what I sketch here is
\emph{certainly} not anti-realism.

I am not against asking questions that, in traditional terms, seem to
bear on issues of realism and instrumentalism.  I am against the focus
on the questions as meaningful and valuable in themselves, without
regard to the roles they may or may not play in the ongoing enterprise
of our scientific attempts to comprehend the physical world.  That
focus, it seems to me, leads only to a sterile form of ideological
back-and-forth that has all but crowded out the possibility of
formulating and addressing questions of real scientific and
philosophical clarity and value.  I take that to be the thrust of the
epigraph from Maxwell at the head of this paper.

\addcontentsline{toc}{section}{\hspace*{-1.3em}\numberline{}References}

\end{document}